\documentclass[letterpaper,showpacs,aps]{revtex4-2}

\usepackage[utf8]{inputenc}
\usepackage[english]{babel}

\usepackage{latexsym}
\usepackage{amssymb}
\usepackage{amsmath}
\usepackage{bm} 
\usepackage{tabularx}
\usepackage{graphicx}
\usepackage{epsfig}

\graphicspath{{images/}}

\usepackage{color}

\begin{document}


\title{Proca stars in excited states}

\author{Carlos Joaquin}
\email{jose.joaquin@correo.nucleares.unam.mx}

\author{Miguel Alcubierre}
\email{malcubi@nucleares.unam.mx}

\affiliation{Instituto de Ciencias Nucleares, Universidad Nacional
Aut\'onoma de M\'exico, Circuito exterior C.U., A.P. 70-543, Ciudad de M\'exico 04510, M\'exico.}


\date{\today}


\begin{abstract}
In this paper, we consider families of solutions for excited states of Proca stars in spherical symmetry. We focus on the first two excited configurations and perform a series of fully non-linear dynamical simulations in order to study their properties and stability. Our analysis reveals that excited Proca stars are always unstable against even very small perturbations, and their dynamical evolution can lead to three different final states: collapse to a black hole, dissipation, or migration to a different configuration in the ground state. We find that migration to the ground state can only occur in a small region of the parameter space of solutions with negative binding energy.
\end{abstract}




\maketitle


\section{Introduction}\label{sec01:intro}

Exotic stars are hypothetical self-gravitating compact objects composed of different types of exotic matter. One particular type of exotic compact object (ECO) is the bosonic soliton, which represents stationary solutions of the Einstein field equations minimally coupled to massive scalar or vector complex fields. The scalar case corresponds to the usual boson stars~\cite{Kaup68,Ruffini69}, while the vector case corresponds to the so-called Proca stars~\cite{BRITO2016291}; these models are commonly referred to as bosonic stars.

There exists a variety of subtypes of bosonic star models that include self-interaction terms, which provide additional parameters to better resemble astrophysical systems; see~\cite{Liebling_2023,Schunck_2003} for reviews. In addition to bosonic stars, there are also oscillatons~\cite{Seidel91} composed of real scalar fields, and Dirac stars~\cite{Finster:1998ws} composed of complex spinor fields, further expanding the array of potential ECOs proposed in the literature. For a comparative analysis and review of these types of objects, see~\cite{Liebling:2012fv,Herdeiro:2020jzx}.

ECOs are theorized to be detectable because of their gravitational interactions with surrounding matter. Extensive research into these compact objects has explored their physical properties, stability, and dynamical behaviors, making them promising candidates for the dark matter component of the Universe. Their ability to clump together and form compact structures, combined with their versatility shaped by the choice of different matter fields and self-interaction potentials, allows them to mimic a variety of astrophysical objects and phenomena. Recently, bosonic stars have been modeled in various astrophysical contexts, such as bosonic dark matter in galactic halos to address the rotation curve problem for galaxies~\cite{Mourelle_2024}. Other studies have analyzed lensing imagescreated by shadows cast in the presence of accretion disks~\cite{Sengo_2024,Rosa_2022}, as well as simulations of vector/boson star mergers and the prediction of their gravitational wave signals in order to compare them with the LIGO/Virgo catalog~\cite{Sanchis-Gual_2019,Bustillo_2023}. For a more detailed discussion, we direct the reader to references~\cite{Brito:2015yfh,Visinelli:2021uve}. All this research offers valuable insights into the nature of the dark sector of the Universe.

In this work, we shall focus on Proca stars, that is, self-gravitating configurations of a complex massive vector field (the Proca field) in four spacetime dimensions, and carrying a conserved Noether charge associated with a global U(1) symmetry. The first such self-gravitating solutions composed of massive spin-1 particles were reported in~\cite{BRITO2016291}. Numerical dynamical evolutions of these configurations have been performed in~\cite{DiGiovanni:2018bvo, Sanchis-Gual:2017bhw}, studying their stability and the various different phenomena that occur under perturbations of the Proca field in the ground state. Additionally, Proca stars have been set up in different astrophysical scenarios, such as those with rotation~\cite{Zhang:2023rwc}, charge~\cite{SalazarLandea:2016bys}, or more generalized models~\cite{Lazarte:2024jyr}. 

In this paper, we focus on excited states of Proca stars, which are significant because they may represent intermediate states during the formation of ECOs. These excited states exhibit distinct properties when compared to the ground state, and have been extensively studied in the case of scalar boson stars in both spherical and axial symmetry~\cite{Balakrishna_1998,Collodel_2017,Brito:2023fwr}. The primary objective of our work is to extend these studies to the case of excited Proca stars, concentrating particularly on their stability under perturbations.

We should emphasize that in this work we will only consider the spherically symmetric case.  In particular, the ground state will be defined as that spherical configuration for which the radial component of the Proca vector potential has no nodes, or alternatively as the spherical state with minimum total mass for a given central value of the Proca scalar potential, while excited states will correspond to solutions with nodes (see sections below for details). It is important to mention here that it was recently shown in~\cite{Herdeiro:2024a} that this spherically symmetric ``ground state" is not stable against non-spherical perturbations, and it does correspond to the minimum mass configuration which is in fact non-spherical. However, here we will only consider spherically symmetric configurations and perturbations.

This paper is organized as follows. In Section~\ref{sec02:model} we present the Einstein–Proca system. In Section~\ref{sec03:PSs} we derive the spherically symmetric field equations for the stationary configurations corresponding to Proca stars and also discuss the total mass, total bosonic charge, and binding energy of the system, as well as the boundary conditions and the numerical methods that we have used. Section~\ref{sec04:results} presents our results for the families of stationary configurations in the first two excited states, as well as the results of their dynamical evolutions under perturbations. We conclude in Section~\ref{sec05:conclusions}.


\section{The Einstein-Proca Model}\label{sec02:model}


\subsection{Action and fields equations}

The Proca field is a massive complex vector field described by the complex potential 1-form $X_{\mu}$, from which we can obtain the field strength 2-form as $W_{\mu \nu} := \nabla_{\mu} X_{\nu} - \nabla_{\nu} X_{\mu}$, where here $\nabla_\mu$ is the usual covariant derivative. The Lagrangian for the Proca field can then be written as:
\begin{equation}
\mathcal{L}_{P} = - \frac{1}{8\pi}  \left( W_{\mu \nu} \bar{W}^{\mu \nu}
+ 2 m^{2} X_{\mu}\bar{X}^{\mu} \right) \; ,
\label{LagrangianProcaField}
\end{equation}
where $\bar{X}$, $\bar{W}$ denote complex conjugates, and $m$ is the mass parameter of the Proca field.

We will consider a self-gravitating Proca field minimally coupled to gravity. The combined Einstein-Proca system is then described by the following action:
\begin{equation}
\mathcal{S} = \int \left(\frac{R}{16\pi} + \mathcal{L}_{P}\right)
\sqrt{-g} \: d^{4}x\; ,
\label{ActionEinsteinProcaModel}
\end{equation}
where $g$ is the determinant of the spacetime metric and $R$ is the curvature scalar. We use units such that $G = c =\hbar = 1$, and the signature of spacetime metric is taken to be $(-,+,+,+)$.  Variation of the action~\eqref{ActionEinsteinProcaModel} with respect to the metric tensor leads to the Einstein field equations:
\begin{equation}
G_{\mu \nu} := R_{\mu \nu} - \frac{1}{2}g_{\mu \nu}R = 8\pi T_{\mu \nu} \; , \label{EinsteinFieldEqs}
\end{equation}
where $G_{\mu \nu}$ is the Einstein tensor for our curved spacetime and $T_{\mu \nu}$ the stress-energy tensor associated with the Proca field, which is derived from the Lagrangian~\eqref{LagrangianProcaField} and takes the form:
\begin{equation}
T_{\mu \nu} = \frac{1}{4\pi} \left[	- W_{\lambda(\mu} \bar{W}_{\nu)}^{\lambda}
- \frac{g_{\mu \nu}}{4} \: W_{\alpha \beta} \bar{W}^{\alpha \beta}
+ m^2 \left( X_{(\mu} \bar{X}_{\nu)} - \frac{g_{\mu \nu}}{2} X_{\lambda} \bar{X}^{ \lambda} \right) \right] \; .
\label{ProcaEnergyTensor}
\end{equation}

On the other hand, variation of the action with respect to the Proca field variables yields the Proca field equations:
\begin{equation}
\nabla_{\mu} W^{\mu \nu} = m^{2} X^{\nu} \; .
\label{ProcaFieldEqs}
\end{equation}
Unlike the case of the real massless spin-1 electromagnetic field, massive spin-1 fields such as the complex Proca field do not exhibit the same gauge invariance. Specifically, in the case of the Proca field the Lorenz condition is no longer a simple gauge choice but becomes instead a dynamical requirement that can be derived directly from~\eqref{ProcaFieldEqs}:
\begin{equation}
\nabla_{\mu} X^{\mu} = 0 \; .
\label{ProcaLorentzCondit}
\end{equation}

It is easy to see that the Einstein--Proca action above is invariant under a global $U(1)$ transformation of the form $X_{\mu} \longrightarrow e^{i\alpha} X_{\mu}$, with $\alpha$ constant. Noether's theorem then implies the existence of a conserved 4-current given by:
\begin{equation}
j^{\mu} := \frac{i}{8\pi} \left[ \bar{W}^{\mu \nu} X_{\nu}
- W^{\mu \nu} \bar{X}_{\nu} \right] \; ,
\label{FourCurrent}
\end{equation}
such that:
\begin{equation}
\nabla_{\mu} j^{\mu} = 0.
\label{DivergencFourCurrent}
\end{equation}


\subsection{The Proca equations in 3+1 formalism}

We work within the framework of the 3+1 formalism of general relativity, where the spacetime ($\mathcal{M}$, $g_{\mu \nu}$) can be foliated by Cauchy hypersurfaces $\Sigma_{t}$ parametrized by a global time function $t$, resulting in the standard Arnowitt--Deser--Misner (ADM) equations (see for example~\cite{Alcubierre08a}). Additionally, by introducing auxiliary variables, one can modify the ADM equations to construct another formalism that is better adapted to stable and long-term numerical simulations, known as the Baumgarte--Shapiro--Shibata--Nakamura (BSSN) formalism~\cite{Shibata95,Baumgarte:1998te}. In particular, this BSSN formalism is the one used in our time evolution code adapted to spherical symmetry~\cite{Alcubierre:2010is}.

The 3+1 split of the metric takes the following form:
\begin{equation}
ds^2 = \left( - \alpha^2 dt^2 + \beta_i \beta^i \right) dt^2 + 2\beta_i dtdx^i
+ \gamma_{ij} dx^i dx^j \;, \label{GeneralLineElement}
\end{equation}
where $\gamma_{ij}$ is the three dimensional metric defined within the spatial hypersurfaces, $\alpha$ is the lapse function, and $\beta^i$ is the shift vector. The unit normal vector $n^{\mu}$ to the spatial hypersurfaces has components given by
\begin{equation}
n^\mu = \left( 1/ \alpha, -\beta^{i}/\alpha\right) \; , \qquad
n_\mu = \left(  -\alpha, 0,0,0 \right) \; .
\label{VectorEuler}
\end{equation}
This timelike unit vector is identified with the 4-velocity of the so-called Eulerian observers.

We can do a 3+1 decomposition of the Proca field by defining the scalar potential $\phi$ and the 3-dimensional vector potential $a_{\mu}$ as:
\begin{equation}
\phi := -n^{\mu} X_{\mu} \; , \qquad
a_\mu := \gamma_{\mu}^{\nu} X_{\nu} \; ,
\end{equation}
where $\gamma_\mu^\nu = \delta_\mu^\nu + n_\mu n^\nu$ is the projector operator onto the spatial hypersurfaces. From these potentials we can reconstruct the Proca potential $X_\mu$ as:
\begin{equation}
X_\mu = a_\mu + n_\mu \phi \; ,
\label{ProcaField}
\end{equation}
Notice that for the vector potential $a^\mu$ we have $a^0=0$, so that from now on we will only consider the spatial components $a^i$, with $a_i := \gamma_{ij} a^j$.  We further define the ``electric" and ``magnetic" fields associated with the Proca field as:
\begin{align}
\mathcal{E}_{\mu} := n^{\nu} W_{\mu \nu} \; , \qquad
\mathcal{B}_{\mu} := n^{\nu} W^{*}_{\mu \nu} \; ,
\end{align}
where $W^{*}_{\mu \nu}$ is the dual field tensor defined as:
\begin{equation}
W^{*}_{\alpha \beta} := - \frac{1}{2} \: E_{\alpha \beta \mu \nu} W^{\mu \nu} \; ,
\end{equation}
with $E_{\alpha \beta \mu \nu}$ the Levi--Civita tensor defined in terms of the totally anti-symmetric symbol $\epsilon_{\alpha \beta \mu \nu}$ as \mbox{$E_{\alpha \beta \mu \nu} := \left|
g \right|^{1/2} \epsilon_{\alpha \beta \mu \nu}$}, with $g$ the determinant of the spacetime metric $g_{\mu \nu}$ (we use a convention such that $\epsilon_{0123} = 1 = - \epsilon^{0123}$). Notice that from the above definitions one can show that the electric and magnetic fields are purely spatial, that is, $n^\mu \mathcal{E}_\mu = n^\mu \mathcal{B}_\mu = 0$.

In terms of the electric and magnetic fields, the Proca field tensor and its dual can now be expressed as:
\begin{align}
W_{\mu \nu} &= n_\mu {\cal E}_\nu - n_\nu {\cal E}_\mu
+ E_{\mu \nu \lambda} {\cal B}^\lambda \; , \label{eq:W-EB} \\
W^*_{\mu \nu} &= n_\mu {\cal B}_\nu - n_\nu {\cal B}_\mu
- E_{\mu \nu \lambda} {\cal E}^\lambda \; , \label{eq:WD-EB}
\end{align}
where the three-dimensional Levi--Civita tensor is now defined as $E_{\mu \nu \lambda} := n^\alpha E_{\alpha \mu \nu \lambda}$.

Based on the previous definitions, and manipulating the Proca field equations~\eqref{ProcaFieldEqs} together with the Lorenz condition~\eqref{ProcaLorentzCondit}, we can obtain the following set of time evolution equations for the Proca potentials and fields:
\begin{align}
\frac{d}{dt} \: \phi &= \alpha \phi K - D_{i} (\alpha a^{i}) \; ,
\label{EvolutionEqScalarPotential} \\
\frac{d}{dt} \: a_{i} &= - \alpha \mathcal{E}_{i} - \partial_{i} (\alpha \phi) \; ,
\label{EvolutionEqVectorPotential} \\
\frac{d}{dt} \: \mathcal{E}^{i}  &=  \alpha \left(	K \mathcal{E}^i + m^{2} a^{i}\right)
+ \left[ D \times (\alpha \mathcal{B})\right]^{i} \; ,
\label{EvolutionEqElectricField}\\
\frac{d}{dt} \: \mathcal{B}^{i} \: &= \alpha K \mathcal{B}^{i}
- \left[ D \times (\alpha \mathcal{E})\right]^{i} \; , \label{EvolutionEqMagneticField}
\end{align} 
with $d/dt := \partial_{t} - \pounds_{\vec{\beta}}$ with $\pounds_{\vec{\beta}}$ the Lie derivative with respect to the shift vector, and where $D_{i}$ is the three-dimensional covariant derivative compatible with the spatial metric $\gamma_{ij}$, and $K$ is the trace of the extrinsic curvature tensor $K_{ij}$. Notice that again we only consider the spatial components of the electric and magnetic fields.

Furthermore, we also obtain the Gauss~\eqref{GaussConstraint} and magnetic~\eqref{MagneticConstraint} constraints:
\begin{align}
D_i \mathcal{E}^{i} + m^{2} \phi = 0 \; , \label{GaussConstraint} \\
D_i \mathcal{B}^{i} = 0 \; . \label{MagneticConstraint}
\end{align}

One can express the 3+1 matter terms coming from the stress--energy tensor~\eqref{ProcaEnergyTensor} in terms of the 3+1 quantities associated with the Proca field.  For the energy density we find:
\begin{equation}
\rho := n^\mu n^\nu T_{\mu \nu}
= \frac{1}{8 \pi} \left[ {\cal E}^2 + {\cal B}^2
+ m^2 \left( \phi^2 + a^2 \right) \right] \; ,
\label{eq:energydensity}
\end{equation}
where:
\begin{equation}
{\cal E}^2 := {\cal E}_\alpha \bar{\cal E}^\alpha = {\cal E}_i \bar{\cal E}^i \; , \quad
{\cal B}^2: = {\cal B}_\alpha \bar{\cal B}^\alpha = {\cal B}_i \bar{\cal B}^i \; , \quad
\phi^2 := \phi \bar{\phi} \; , \quad
a^2 := a_i \bar{a}^i \; .
\end{equation}
For the momentum density we have:
\begin{equation}
J^i := - \gamma^{i \mu} n^\nu T_{\mu \nu}
= \frac{1}{8 \pi} \left[ E^i{}_{jk} \bar{{\cal E}}^j {\cal B}^k
+ m^2 a^i \bar{\phi} + {\rm c.c.} \right] \; ,
\label{eq:momentumdensity}
\end{equation}
where c.c. indicates the complex conjugate of the previous terms.
Finally, for the 3D stress tensor we have:
\begin{align}
S_{ij} := \gamma^\mu_i \gamma^\nu_j T_{\mu \nu}
&= \frac{1}{8 \pi} \left\{ \gamma_{ij} \left( {\cal E}^2 + {\cal B}^2 \right)
- \left( {\cal B}_i \bar{\cal B}_j + {\cal E}_i \bar{\cal E}_j + {\rm c.c.} \right) \right. \nonumber \\
&+ \left. m^2 \left[ \left( a_i \bar{a}_j + {\rm c.c.} \right)
- \gamma_{ij} \left( a^2 - \phi^2 \right) \right] \right\} \; .
\label{eq:stresstensor}
\end{align}

\vspace{5mm}

One can also find 3+1 expressions for the conserved Noether current.  In particular, the particle density
is given by:
\begin{equation}
\rho_Q := - n_\mu j^\mu = \alpha j^0
= \frac{i}{8 \pi} \left[ a_k \bar{\cal E}^k - {\rm c.c.} \right] \; ,
\label{eq:chargedensity}
\end{equation}
while for the particle flux we find:
\begin{equation}
j^i_Q := \gamma^i_\mu j^\mu
= \frac{i}{8 \pi} \left[ \phi \bar{\cal E}^i
+ E^{ijk} a_j \bar{\cal B}_k - {\rm c.c.} \right] \; .
\label{eq:particleflux}
\end{equation}


\section{Proca stars}
\label{sec03:PSs}

To study stationary configurations corresponding to Proca stars we first assume spherical symmetry. The line element in spherical coordinates $(t, r,\theta, \phi)$ can then be written as:
\begin{equation}
ds^2 = - \alpha^2 dt^2 + \psi^4 \left( A dr^2 + r^2 B d \Omega^2 \right) \; , \label{LineElementFormalism}
\end{equation}
with $\psi$ a conformal factor, and where $d \Omega^2 = d \theta^2 + \sin^2 \theta d \phi^2$ is the standard solid angle element. In general, the metric functions $(\alpha,\psi,A,B)$ are functions of both time $t$ and radius $r$, though for the case of stationary configurations they will depend only on $r$. Notice that in principle we do not need to introduce the conformal factor $\psi$, and indeed we will take it to be equal to 1 for our stationary configurations, but we do need it for the BSSN formalism used in the dynamical simulations that we will consider later. In the context of this work, the assumption that the shift vector vanishes, $\beta^{i} = 0$, is adequate for both stationary configurations and our dynamical evolutions.  In order to obtain our stationary configurations, we will further restrict ourselves to using the areal radial coordinate, so in this section we will take $B=\psi=1$.

Proca stars (PSs) are self-gravitating solutions for which the Proca field exhibits a harmonic time dependence, resulting in a time-independent stress-energy tensor which leads to a static spacetime metric. As mentioned above, here we will only consider the case of spherical symmetry. Notice that in spherical symmetry the magnetic field vanishes, while the vector potential and the electric field only have a non-zero radial component. We then propose the following ansatz for the scalar potential and the radial components of the vector potential and the electric field:
\begin{align}
\begin{split}
\phi(r,t) &= \varphi(r) e^{-i \omega t} \; , \\
a_r (r,t) &= \textit{i} a(r) e^{-i \omega t} \;, \\
\mathcal{E}^r (r,t) &= e(r) e^{-i \omega t} \;, \label{Ansantzs}
\end{split}
\end{align}
with $\omega$ a real constant corresponding to the oscillation frequency of the Proca field, and where the functions $\varphi(r)$, $a(r)$ and $e(r)$ are real and depend only on the radial coordinate $r$. The explicit factor of $i$ introduced in the ansatz for the radial component of the vector potential $a_r$ is there to guarantee that both the momentum density~\eqref{eq:momentumdensity} and the particle flux coming from the conserved Noether current~\eqref{eq:particleflux} vanish, as they should for a static solution.  In our case the momentum density reduces to:
\begin{equation}
J_r = \frac{m^2}{8 \pi} \left[ a_r \bar{\phi} + c.c \right]
= \frac{m^2}{8 \pi} \left[ i a \varphi + c.c \right]
= 0 \; ,
\end{equation}
while the particle flux reduces to:
\begin{equation}
j^r_Q = \frac{i}{8 \pi} \left[ \phi \bar{\mathcal{E}}^r - c.c. \right]
= \frac{i}{8 \pi} \left[ \varphi e - c.c. \right]
= 0 \; .
\end{equation}

By substituting our harmonic anzats~\eqref{Ansantzs} into the evolution equations~\eqref{EvolutionEqScalarPotential}---\eqref{EvolutionEqElectricField} and the Gauss constraint~\eqref{GaussConstraint}, we can derive a set of ordinary differential equations in $r$ to solve for our Proca star configurations. In order to simplify some expressions, we work with $F(r):= \alpha \varphi$ instead of $\varphi$ itself.  Once the system is solved, $\varphi$ can be immediately reconstructed as $\varphi = F/\alpha$.

From the evolution equation for $a_r$, equation~\eqref{EvolutionEqVectorPotential}, we find:
\begin{equation}
\partial_r F = - \omega a - \alpha e A \; .
\end{equation}
On the other hand, from the evolution equation for $\mathcal{E}^r$, equation~\eqref{EvolutionEqElectricField}, we can solve explicitly for $e$ to find:
\begin{equation}
e = - \frac{\alpha m^2 a}{\omega A} \; ,
\label{eq:e-solution}
\end{equation}
so that our equation for $F$ then becomes:
\begin{equation}
\partial_r F = \omega a \left(\frac{\alpha^2 m^2}{\omega^2}	- 1 \right) \; .
\label{EDO-F}
\end{equation}

We can also obtain a differential equation for $a$ from the evolution equation for $\phi$, equation~\eqref{EvolutionEqScalarPotential}. We find:
\begin{equation}
\partial_r a = \frac{\omega \varphi A}{\alpha} - a \left( \frac{2}{r} - \frac{\partial_r A}{2A}
+ \frac{\partial_r \alpha}{\alpha} \right) \; .
\label{eq:aprime}
\end{equation}

Finally, from the Gauss constraint we find:
\begin{equation}
\partial_r e = - e \left( \frac{2}{r} + \frac{\partial_r A}{2A} \right) - m^2 \varphi \; .
\label{eq:eprime}
\end{equation}
Notice that this last equation in principle is not necessary, as we already have an explicit expression for $e$ in equation~\eqref{eq:e-solution} above.  Indeed, one can show that this last equation is completely consistent with that solution.  However, equation~\eqref{eq:eprime} will be necessary when we consider perturbed Proca stars below.

We still need equations for the lapse function $\alpha$ and the radial metric $A$.
For the radial metric we use the Hamiltonian constraint, which in spherical symmetry
and with our coordinate choices reduces to:
\begin{equation}
\partial_r A = A\left[\frac{(1-A)}{r} + 8 \pi r A \rho \right] \; .
\label{EDO-A}
\end{equation}
The equation for the lapse comes from our choice of using the areal radius, which in turn implies that the angular component of the extrinsic curvature has to vanish for all time, which through the ADM equations leads to the so-called ``polar-areal" gauge condition that takes the form:
\begin{align}
\partial_r \alpha = \alpha \left[	\frac{\left(A - 1\right)}{2r} + 4 \pi r A S^r_r \right] \; .
\label{EDO-alpha}
\end{align}

Notice that the equations for $\alpha$ and $A$ above have matter terms on the right-hand side corresponding to the energy density $\rho$ and the component $S^r_r$ of the stress tensor. Using now equations~\eqref{eq:energydensity} and~\eqref{eq:stresstensor} one can show that in our case those quantities reduce to:
\begin{align}
\rho  &= + \frac{1}{8 \pi} \left[ A e^2 + m^2 \left( \varphi^2 + a^2 / A \right) \right] \; .
\label{eq:rho-proca} \\
S^r_r &= - \frac{1}{8 \pi} \left[ A e^2 - m^2 \left( \varphi^2 + a^2 / A \right) \right] \; .
\label{eq:Srr-proca}
\end{align}

As a final comment, equation~\eqref{eq:aprime} above has terms on the right-hand side that involve derivatives of $\alpha$ and $A$. But we can substitute those derivatives using the equations for $\alpha$ and $A$ that we just found, so that the equation for $a$ now becomes:
\begin{equation}
\partial_r a = \frac{A\omega F}{\alpha^2} - a \left[	\frac{1}{r} \left(1 + A	\right)
+ 4 \pi r A \left(S^r_r - 2 \rho \right) \right] \; .
\label{EDO-a}    
\end{equation}

The final system of equations that we need to solve for ($A,\alpha,F,a$) is then~\eqref{EDO-F}, \eqref{EDO-A}, \eqref{EDO-alpha} and~\eqref{EDO-a}, with $e$ given by~\eqref{eq:e-solution}.


\subsection{Boundary conditions}

We need to solve the system of first-order differential equations \eqref{EDO-F}, \eqref{EDO-A}, \eqref{EDO-alpha}, \eqref{EDO-a} for the functions $(A,\alpha,F,a)$ in order to obtain the family of Proca stars, subject to a series of physical boundary conditions.  The boundary conditions for the radial metric come from demanding regularity at the origin:
\begin{equation}
A(r=0) = 1 \; , \qquad \left. \partial_r A \right|_{r=0} = 0 \; .
\end{equation}
Notice that we must ask for $A(r=0)=1$ in order for the metric to be locally flat there.
Also, we must have $\lim_{r \rightarrow \infty }A=1$ in order to have an asymptotically flat spacetime, though the structure of equation~\eqref{EDO-A} guarantees that this is the case if the energy density decays fast enough.  For the lapse we ask for regularity at the origin, and also to recover Minkowski spacetime far away, which corresponds to:
\begin{equation}
\lim_{r \rightarrow \infty} \alpha = 1 \; , \qquad
\left. \partial_r \alpha \right|_{r=0} = 0 \; .
\end{equation}

For the scalar potential $\varphi$ (or equivalently $F$) we ask again regularity at the origin, and $\varphi$ going to zero at infinity:
\begin{equation}
\lim_{r \rightarrow \infty} \varphi = 0 \; , \qquad
\left. \partial_r \varphi \right|_{r=0} = 0 \; .
\end{equation}

Finally, the vector potential $a$ must also go to zero at infinity, and since it is the radial component of a vector it must vanish at the origin as well:
\begin{equation}
a(r=0) = 0 \; , \qquad
\lim_{r \rightarrow \infty} a = 0 \; .
\end{equation}

\vspace{5mm}

We can in fact be somewhat more precise.  For an asymptotically flat spacetime, for large $r$ the equations for $\varphi$ and $a$ become (taking $A=1$ and $\rho=S^r_r \simeq 0$):
\begin{equation}
\partial_r \varphi = \frac{\omega a}{\alpha_\infty}
\left( \frac{\alpha^2_\infty m^2}{\omega^2} - 1 \right) \; , \qquad
\partial_r a = \frac{\omega \varphi}{\alpha_\infty} \; ,
\end{equation}
with $\alpha_\infty$ the asymptotic value of the lapse (which as we already mentioned should be equal to 1, but see below). Combining both equations we find:
\begin{equation}
\partial^2_r  \varphi = \left( m^2 - \frac{\omega^2}{\alpha^2_\infty} \right) \varphi \; ,
\end{equation}
which can be easily solved to find:
\begin{equation}
\varphi = e^{\pm (m^2 - \omega^2/\alpha^2_\infty)^{1/2} r} \; .
\end{equation}
From this we see that in order for $\varphi$ to be real we must necessarily have \mbox{$\omega^2<\alpha^2_\infty m^2$}.  But notice that in that case we can have both increasing and decreasing exponential solutions.  The increasing solutions are unphysical, as they are not compatible with an asymptotically flat spacetime.  In fact, one finds that we only have exponentially decaying solutions for a discrete set of values for the frequency $\omega$, which implies that we need to solve an eigenvalue problem.

\vspace{5mm}

We can also give some more detail for the boundary conditions at the origin. The equation for $\partial_r a$ has a term that goes as $a/r$.  This implies that for regularity close to the origin we must have $a \sim k_a r$, with $k_a$ some constant. Substituting this in the equation
we find:
\begin{equation}
k_a = \frac{\omega \varphi_0}{3 \alpha_0} \; ,
\end{equation}
with $\alpha_0=\alpha(r=0)$ and $\varphi_0=\varphi(r=0)$. In the same way,
we see that in the equation for $\partial_r A$ there is a term that goes as
$(1-A)/r$. This means that close to the origin we must have $A \sim 1 + k_A r^2$.  Substituting this in the equation we now find:
\begin{equation}
k_A = \frac{8 \pi}{3} \: \rho_0 \; ,
\end{equation}
with $\rho_0 = \rho(r=0)$. These conditions turn out to be very useful
in order to start the numerical integration of our equations from the origin.

\vspace{5mm}

Numerically, we solve our system of equations starting from the origin, where
we choose as our free parameter the value of $\varphi_0$ and a trial value for
the frequency $\omega$.  We then integrate the system outward (using a fourth-order
Runge--Kutta method) and keep modifying the value of $\omega$ with a shooting algorithm until we find a solution that decays far away.

There is one last detail that must be considered. Since initially we do not know the 
value of the lapse $\alpha$ at the origin we start by taking $\alpha_0=1$.  However, what we really want is $\alpha$ to become zero at infinity. But this is not a serious problem since the equation for $\partial_r \alpha$ is linear in $\alpha$, so we can always rescale the solution at the end using the asymptotic value of the lapse $\alpha_\infty$ (obtained by extrapolation assuming that the lapse far away behaves as $\alpha \simeq \alpha_\infty - c/r$ for some positive constant $c$). However, in order for this to still be a solution of the whole system we must also rescale the frequency $\omega$ by the same factor. In particular, this implies that our initial trial frequency is in general such that $\omega>m$, but once we rescale it we will have $\omega < m$.  Notice also that if we solve the system for $F$ instead of $\varphi$, then we also need to rescale $F$ by the same factor to maintain $\varphi=F/\alpha$ the same as before.


\subsection{Total mass and particle number}

When working in terms of the areal radius, the total mass $M$ of the spacetime can be shown to be the integral of the energy density over a flat volume element (this can be shown directly from the Hamiltonian constraint for a static spacetime when using the areal radius):
\begin{align}
M = 4 \pi \int_{0}^{\infty} \rho r^2 dr \; ,
\label{IntegratedMass}
\end{align}
with the energy density of matter $\rho$ given by equation~\eqref{eq:rho-proca}.

Similarly, the existence of a Noether current implies a conserved charge $Q$ given by:
\begin{equation}
Q := \int \rho_Q \gamma^{1/2} d^3 x = 4 \pi \int_{0}^{\infty} \rho_Q A^{1/2} r^2 dr \; ,
\label{NoetherCharge}
\end{equation}
with $\rho_Q$ the particle density given by~\eqref{eq:chargedensity}, and where we used the fact that the determinant of the spatial metric is given by $\gamma = A r^2 \sin^2 \theta$ (notice that for this integral we do need to use the physical curved volume element). In our case the particle density $\rho_Q$ reduces to:
\begin{equation}
\rho_Q = - \frac{ae}{4 \pi} = \frac{\alpha m^2 a^2}{\omega A} \; .
\end{equation}

The integrated charge $Q$ represents the total number of bosonic particles, so that the product $mQ$ corresponds to the total rest mass of the system. On the other hand, the total mass $M$ includes contributions from kinetic and potential energies, as well as the rest mass. Therefore, we can define a binding energy as:
\begin{align}
U := M - mQ \; , \label{BindingEnergy}
\end{align}
in order to isolate the kinetic and potential contributions. This means that those solutions with $mQ > M$ will have a negative binding energy $U<0$ and will correspond to gravitationally bound stars. Conversely, solutions with $mQ < M$ have positive binding energy $U>0$, making them gravitationally unbound.  In the case of scalar boson stars gravitationally unbound configurations have been found to be unstable against perturbations~\cite{Balakrishna:1997ej,Guzman04,Guzman09}, so one might expect the same to be true for Proca stars, and this is indeed what we find in our simulations (see below).  The definition of the binding energy above has been used frequently in the literature on bosonic stars, starting from the early work on boson stars (see e.g.~\cite{Seidel90}). A formal justification for this definition can be found in~\cite{Loginov_2015}.


We can also use the mass integral above to define an effective radius $R_{99}$, which contains 99\% of the total mass. This effective radius is necessary because, as mentioned above, for Proca stars the energy density decays exponentially implying that there is no real surface.


\subsection{Perturbations}

In previous sections we derived the equations that we need to solve in order to obtain stationary Proca star configurations; we will show families of such solutions for the ground state and the first two excited states in Section~\ref{sec:stationary} below. However, one of the aims of this paper is to study the dynamical stability of perturbed Proca star configurations.  We will therefore also consider self-consistent perturbations to the stationary Proca star solutions.

Once we have a stationary solution, we will construct a perturbation to it in the following way:

\begin{enumerate}

\item We add a small perturbation to the scalar potential $\varphi(r)$ (in practice, we add it to $F(r)$), typically a Gaussian, leaving the vector potential $a(r)$ unchanged.  The perturbation in $\varphi$ (or $F$) must be even with respect to reflections on the origin to maintain regularity there. For example, if we add a Gaussian centered on a point $r=r_0$ away from the origin, we must also add a similar Gaussian centered on $r=-r_0$ to guarantee that we still have $\partial_r \varphi(r=0)=0$.  Notice also that if we only modify the real part of the scalar potential while leaving the vector potential purely imaginary, the momentum density $J_r$ remains zero. Similarly, since the scalar potential remains real, the electric field $\mathcal{E}^r$ will also be purely real, so that the particle flux $j^r_Q$ also remains zero.

\item We solve again the Hamiltonian constraint~\eqref{EDO-A}, the polar slicing condition~\eqref{EDO-alpha}, and the Gauss constraint~\eqref{eq:eprime} for the functions $(A,\alpha,e)$.  Notice that we now cannot use the explicit solution for the electric field $e(r)$ given by~\eqref{eq:e-solution} since it is no longer valid for a perturbed star, so we must solve for $e(r)$ using equation~\eqref{eq:eprime}.

\end{enumerate}

The procedure just described results in consistent initial data for a perturbed Proca star. Of course, the perturbed star will no longer be stationary.  However, as mentioned above, the momentum density and particle flux are initially zero, which is consistent with taking an initially vanishing extrinsic curvature.  Our perturbed initial data will then correspond to a moment of time symmetry, that is, the initial time derivative of the spatial metric will vanish, but its second time derivative will not.

Once we have the perturbed initial data we evolve it in time using the OllinSphere code (see below), which implements a BSSN formulation adapted to spherical symmetry~\cite{Alcubierre:2010is}. For the evolution of the Proca field variables we use equations~\eqref{EvolutionEqScalarPotential}-\eqref{EvolutionEqElectricField}, which in our case reduce to:
\begin{align}
\partial_t \phi &= \alpha K \phi - \frac{1}{A \psi^4} \: \partial_r \left( \alpha a_r \right)
+ \frac{\alpha a_r}{A \psi^4} \left( \frac{\partial_r A}{2A} - \frac{\partial_r B}{B}
- 2 \: \frac{\partial_r \psi}{\psi} - \frac{2}{r} \right) \; , \\
\partial_t a_r &= - \alpha A \psi^4 \mathcal{E}^r
- \partial_r \left( \alpha \phi \right) \; , \\
\partial_t \mathcal{E}^r &= \alpha \left( K \mathcal{E}^r + \frac{m^2 a^r}{A \psi^4} \right) \; .
\end{align}

Notice that even if for our initial data we took the conformal factor $\psi$ and the angular metric component $B$ equal to 1 in the metric~\eqref{LineElementFormalism}, this will not remain true for a dynamical simulation.  We must also remember that all three quantities $(\phi,a_r,\mathcal{E}^r)$ are complex, so we must evolve their real and imaginary parts separately.

Finally, as gauge conditions during the evolution we take a vanishing shift vector $\beta^r=0$, and we use the well-known ``1+log" slicing condition which takes the form:
\begin{equation}
\partial_t \alpha = -2 \alpha K \; .
\label{eq:1+log}
\end{equation}
This slicing condition has been shown to be very robust in practice, allowing for stable and long lived dynamical evolutions~\cite{Alcubierre08a,Arbona99,Alcubierre02a,Alcubierre01a}.


\section{Results}
\label{sec04:results}

Before discussing our main results, we will say something about the numerical methods used here.  For finding the stationary solutions, as well as initial data for the perturbed solutions, the system of ordinary differential equations is integrated numerically using a fourth-order Runge-Kutta method. The code takes a central value of the potential scalar $\varphi_0$ as an input parameter, and employs a shooting algorithm to determine the value of the frequency $\omega$ that corresponds to the asymptotically decaying solution of the scalar potential. For simplicity, we set the value of the Proca mass parameter to $m=1$, but notice that the solutions can later be rescaled to arbitrary values of $m$ since our system of equations is invariant under the rescaling:
\begin{equation}
\begin{aligned}
m \rightarrow \lambda m \; , \qquad
\omega \rightarrow \lambda \omega \; , \qquad
r \rightarrow r/\lambda \; , \qquad
e \rightarrow \lambda e \; .
\end{aligned}
\end{equation}

Although we do not show them here, we have performed extensive tests to check that our numerical solutions do indeed converge to fourth order, as expected. In particular, we have checked that the Hamiltonian and Gauss constraints converge to zero.  Typical solutions are obtained with a spatial resolution of $\Delta r=0.01$, with the boundaries located at $r \simeq 100$.  With this resolution, the maximum violations of the Hamiltonian and Gauss constraints are of the order of $10^{-9}$.

For a given value of $\varphi_0$, and a given resolution, we determine the frequency $\omega$ to machine round-off error.  However, the value of $\omega$ does change slightly with resolution.  For example, when changing the resolution from $\Delta r = 0.01$ to $\Delta r = 0.02$, while keeping the same value of $\varphi_0$ and the same position of the boundaries, we typically find changes in the value of the frequency $\omega$ in the 11th significant figure.

\vspace{5mm}

As mentioned above, for the dynamical evolutions we use the OllinSphere code.  This is a general purpose finite difference numerical relativity code in spherical symmetry based on the BSSN formulation, that uses fourth-order finite differencing in space, and a method of lines evolution in time with a fourth-order Runge--Kutta integrator.  This code has been extensively tested for consistency and accuracy in many different scenarios~\cite{Alcubierre:2014joa,Alcubierre:2019qnh,Degollado:2020lsa,Jimenez2022a,Jimenez2022b}. 

For our dynamical simulations we typically use a larger value of the spatial increment, $\Delta r = 0.05$, with a Courant parameter $\Delta t / \Delta r = 0.5$ to ensure numerical stability.  This is to reduce computational demands since the evolutions can be extremely long, with several million time steps, in order to properly see the late-time behavior.   We monitor the constraint violation during our simulations, and typically find that the root mean square (RMS) norm of the constraints grows slowly with time, but converges to zero with increased resolution.  We have also checked that moving the boundaries further away does not affect our results in a significant way.


\subsection{Families of stationary solutions for ground and excited states}
\label{sec:stationary}

In this section we will present the properties of the stationary configurations for three families of Proca star solutions, corresponding to the ground state and the first two excited states. Table~\ref{Tabla-Min-U} shows the parameters that correspond to the maximum mass configuration for all three families. In particular, we show the central value of the scalar potential $\varphi_0$, the corresponding frequency $\omega$, the total mass $M$, the binding energy $U$, as well as the effective radius $R_{99}$ and effective compactness defined as $C_{99} := M/R_{99}$. Our data can be compared with that reported in~\cite{Brito:2015pxa,Sanchis-Gual:2017bhw,Herdeiro:2017fhv} for the ground state. However, we also explore the properties of the first two excited states, which correspond to solutions with one or two nodes in the vector potential $a(r)$ (notice the scalar potential $\varphi(r)$ already has one node even in the ground state, see below).

\begin{table}
\centering
\begin{tabular}{ || m{5em}  m{2cm} m{2cm}  m{2cm}  m{2cm} m{2cm}  m{2cm} ||}
\hline
State & $\varphi_0$ & $\omega$ & $M_{max}$ & $U$ & $C_{99}$ & $R_{99}$ \\ 
\hline\hline 
ground & 0.1410 & 0.8698 & 1.058 & -0.030 & 0.0900 & 11.75 \\ 
\hline 
1° excited & 0.1636 & 0.8828 & 1.797 & -0.047 & 0.0898 & 20.01 \\ 
\hline 
2° excited & 0.1805 & 0.8880 & 2.526 & -0.064 & 0.0899 & 28.07 \\ 
\hline 
\end{tabular}
\caption{We show the parameters that characterize the maximum mass configurations for Proca stars in the ground state and the first two excited states.}
\label{Tabla-Min-U}
\end{table}

\begin{figure}[!htbp]
\centering
\includegraphics[width=1\textwidth]{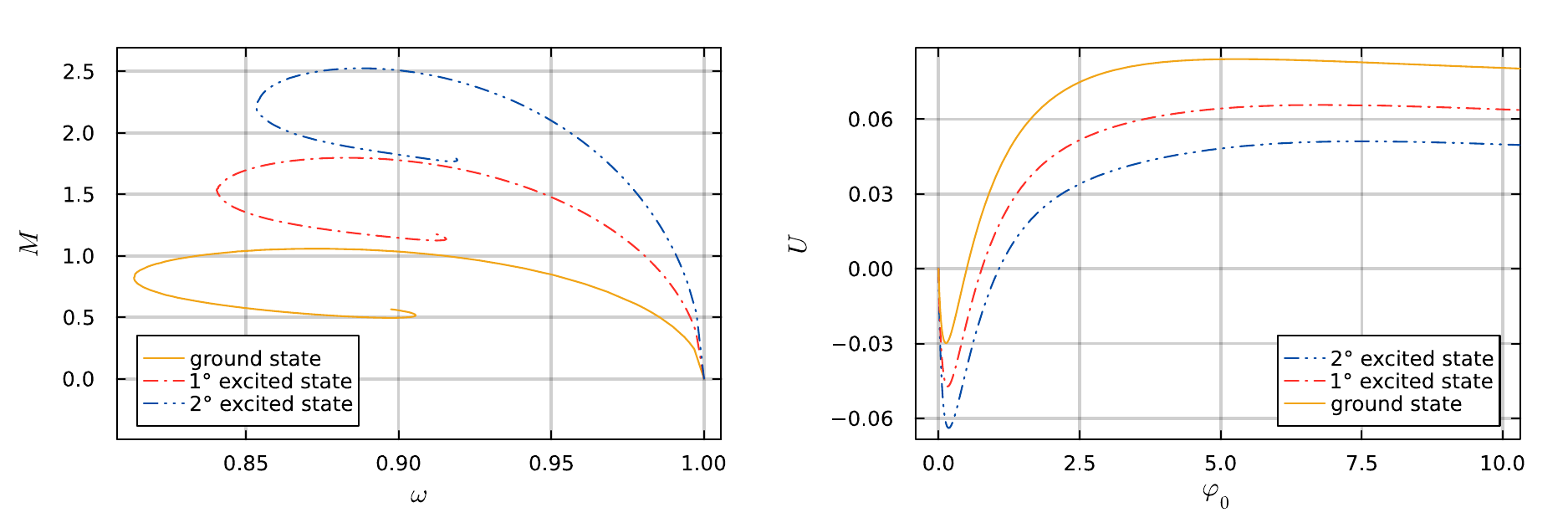}
\caption{\textit{Left panel}: Total integrated mass $M$ (in units of $M^{2}_{Planck}/m$) as a function of frequency $\omega$. We observe that as the excitation level of the star increases the total mass $M$ also increases. \textit{Right panel}: Binding energy $U$ as a function of $\varphi_0$. For each of the families $U$ has a global minimum, which corresponds to the solutions with maximum mass of Table~\ref{Tabla-Min-U}.}
\label{ID_Mass_BE}
\end{figure}

We show our three different families of solutions in Figure~\ref{ID_Mass_BE}.  The left panel shows the total integrated mass $M$ as a function of the oscillation frequency $\omega$, while the right panel shows the binding energy $U$ as a function of the central value of the scalar potential $\varphi_0$.  We can observe that excited states have, in general, higher masses. Additionally, we also find that the frequency domain is bounded for all three families, showing the characteristic spiral pattern in the mass versus frequency plot that also appears for boson stars, with the minimum frequency increasing with the excitation level.  Also, for all three families, there is a value of the central field for which the binding energy reaches a minimum, which in fact corresponds to the solutions with maximum mass.  As we increase the value of $\varphi_0$ further, the binding energy also increases and eventually becomes positive, corresponding to solutions that are no longer gravitationally bound.

Solutions with negative binding energy in all three families can be further separated into solutions to the left and to the right of the minimum binding energy $U$ (maximum mass $M$) in the right panel of Figure~\ref{ID_Mass_BE}. Naively, one could expect that all solutions with negative binding energy should be stable under perturbations since they represent gravitationally bound stars. However, as has already been shown~\cite{Sanchis-Gual:2017bhw}, this is not true even for solutions in the ground state: configurations to the right of the minimum in the $U$ vs. $\phi_0$ plot are unstable to perturbations, and can either collapse to a black hole or migrate to a configuration on the stable left branch (this is the same in the case of boson stars). Having a solution with $U<0$ is therefore a necessary but not sufficient condition for the stability of the Proca stars in the ground state. The purpose of this paper is to investigate whether something similar happens for solutions in excited states. We will return to this point when we consider dynamical simulations below.

\begin{figure}[!htbp]
\centering
\includegraphics[width=1\textwidth]{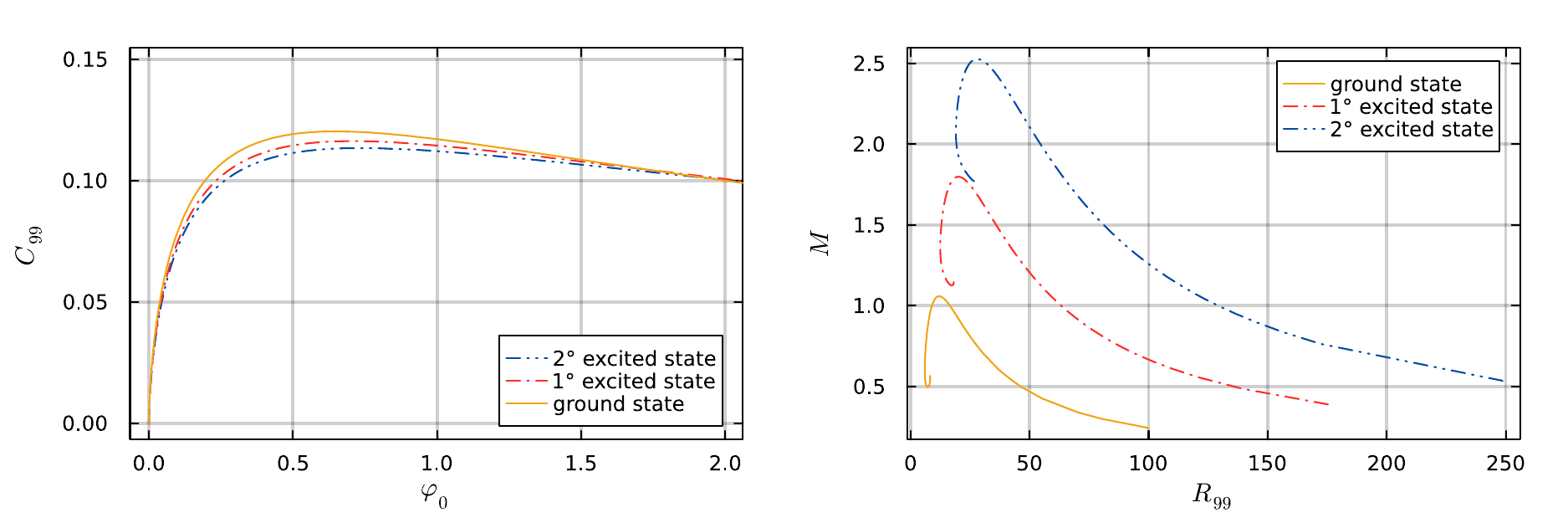}
\caption{\textit{Left panel.}: The effective compactness defined as $C_{99} = M/R_{99}$ versus $\varphi_0$. \textit{Right panel}: Total mass $M$ as a function of the effective radius $R_{99}$. We observe that excited Proca stars increase in both mass and radius.}
\label{ID_C99_Mass_R99}
\end{figure}

In Figure~\ref{ID_C99_Mass_R99} we consider the effective radius and compactness of our three families of solutions. The right panel shows the total mass $M$ as a function of the effective radius $R_{99}$, while the left panel shows the effective compactness defined as $C_{99} = M/R_{99}$ as a function of $\varphi_0$.  Notice that despite the fact that Proca stars in the ground state have in general a lower mass than those in excited states, they exhibit greater compactness. This is not too surprising, since Proca stars in excited states are generally larger in size, as can be seen in the right panel of the figure. For the ground state family the effective compactness reaches a maximum value of $C_{99} \simeq 0.12$.  This value is still far from $C=0.5$ corresponding to the compactness of a Schwarzschild black hole, but is comparable with other kind of compact objects~\cite{Cardoso:2019rvt}.

\begin{figure}[!htbp]
\centering
\includegraphics[width=1\textwidth]{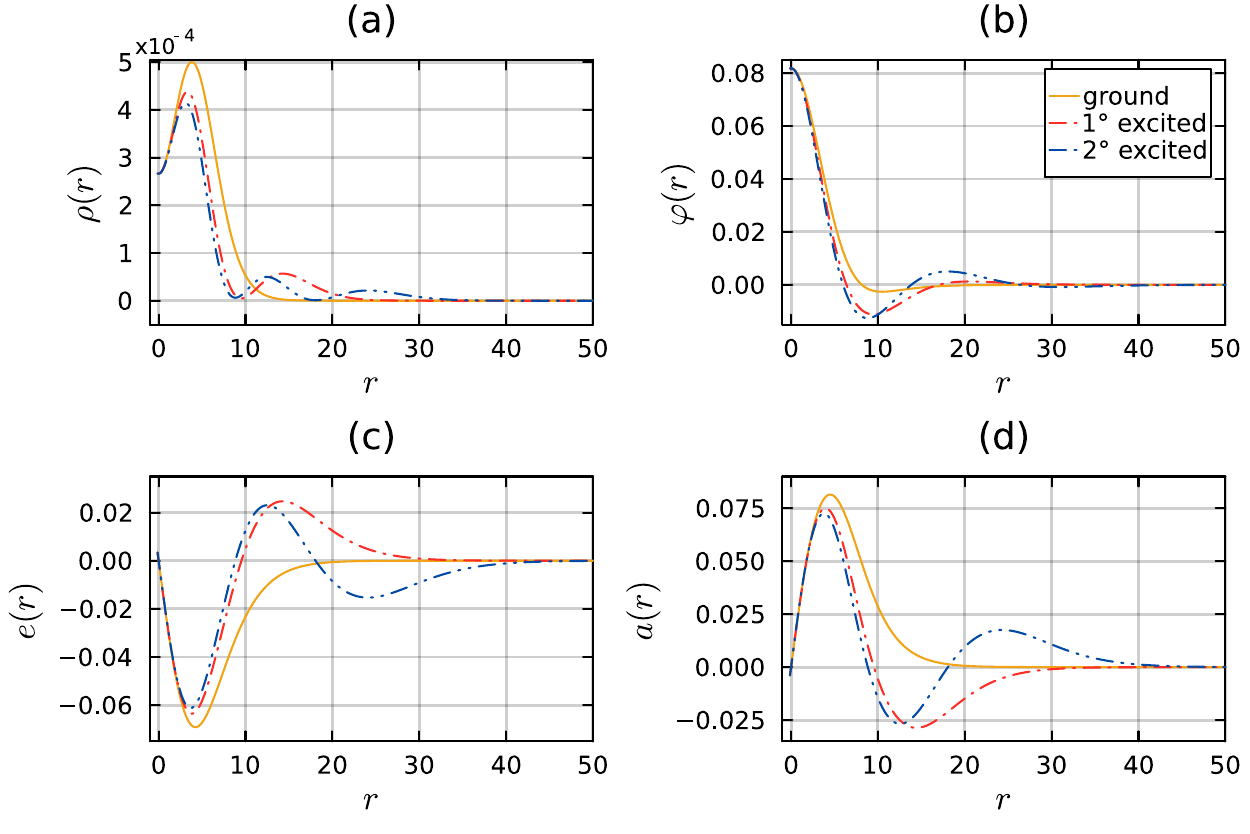}
\caption{We show the solutions for the ground state and the first two excited states for the same central value of the scalar potential $\varphi_0 =0.081$ (the color scheme is the same for all four figures). Panel (a) shows energy density $\rho(r)$.  Panels (b), (c), and (d) show the profiles for the scalar potential $\varphi(r)$, the electric field $e(r)$, and the radial component of vector potential $a(r)$.}
\label{ID_Profile_functions_model}
\end{figure}

In order to better illustrate the properties of our solutions, we consider three particular configurations in Figure~\ref{ID_Profile_functions_model}.  We choose three solutions with the same central value of the scalar potential $\varphi_0 = 0.081$: One in the ground state, one in the first excited state, and one in the second excited state.  All three configurations can be shown to have negative binding energy. Panel (a) shows the density profile $\rho(r)$ for all three solutions.  We notice that the ground state possesses a single core of matter, while the excited states have a core and shells surrounding it, and as a result the effective radii of the excited configurations are larger. The first excited state has one such shell outside the core, while the second excited state has two shells (compare this with the analogue case for excited boson stars in~\cite{Brito:2023fwr}).

The other three panels of the figure show the profile functions $\varphi(r)$, $e(r)$ and $a(r)$ for the ground state and the first two excited states. All these functions decay rapidly far away, as expected.  We can also observe that the number of nodes increases with the excitation levels. The scalar function $\varphi(r)$ in the ground state already has one node, followed by the first and second excited states with two and three nodes, respectively. Additionally, in the ground state both the vector potential $a(r)$ and the Proca electric field $e(r)$ have no nodes, while they have one node in the first excited state and two nodes in the second.


\subsection{Dynamical evolutions and stability}
\label{sec:perturbed}

The values presented in Table~\ref{Tabla-Min-U} will serve as a reference for choosing models for our dynamical evolutions of perturbed Proca star configurations. In all the numerical simulations presented below, we always add a small Gaussian perturbation centered at the origin and with unit width to the scalar potential $\varphi(r)$, while leaving the vector potential $a(r)$ unchanged. The amplitude of this perturbation is always taken as 5\% of the maximum amplitude of the scalar potential of the original stationary solution. As mentioned above, once we add the perturbation to $\varphi(r)$ we solve again the Hamiltonian and Gauss constraints to obtain self-consistent initial data.

We should mention here the fact that we have settled on a 5\% perturbation in order to both be consistent across all simulations, and also to have a large enough perturbation that allows us to see the final stages of the evolution in a reasonable computational time, be it collapse to a black hole, migration, or dispersion. We have checked that smaller perturbations give similar results, but the instabilities take much longer to develop.  Notice that for unstable configurations one does not need to add a perturbation by hand in order to trigger an instability, since numerical truncation error will trigger it on its own given enough time (we have checked that this is indeed the case).  However, this has two disadvantages: First, for high resolutions the numerical truncation error is very small and the instability can take a very long time to develop, and second, triggering a numerical instability with pure truncation error makes it impossible to check for convergence of the late time behavior since at higher resolution the effective perturbation is smaller. Because of this, we prefer to add a finite self-consistent perturbation to our stationary solutions as described above.


\subsubsection{First excited state}

\begin{figure}[!htbp]
\centering
\includegraphics[width=0.6\textwidth]{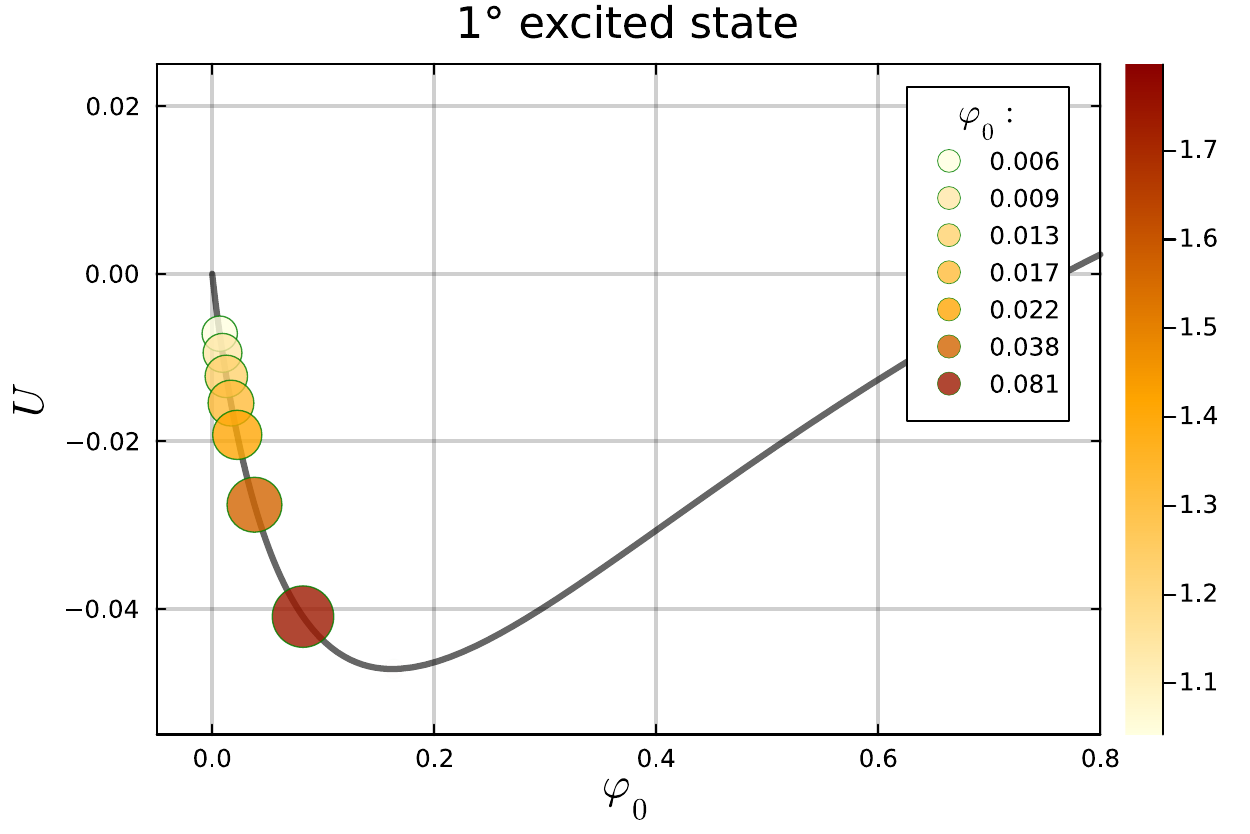}
\caption{Binding energy $U$ as a function of $\varphi_0$ for the first excited state.  The circles show the 7 models that we considered for our perturbed dynamical evolutions. The color bar indicates the scale of the total integrated mass $M$ of the unperturbed stationary configurations. The size of each circle in the plot is proportional to the total mass, and increases as we approach the the minimum of the binding energy.}
\label{FirstState_BE_Models}
\end{figure}

Figure~\ref{FirstState_BE_Models} shows a plot of the binding energy $U$ for Proca stars in the first excited state as a function of the central value of the scalar potential $\varphi_0$. In the figure we also indicate 7 different models that we considered for our dynamical evolutions of perturbed initial data. Notice that all these models have negative binding energy, and are to the left of the minimum in $U$ (corresponding to the maximum mass $M$), the region that one would expect to be more stable.

\begin{figure}[!htbp]
\centering
\includegraphics[width=0.6\textwidth]{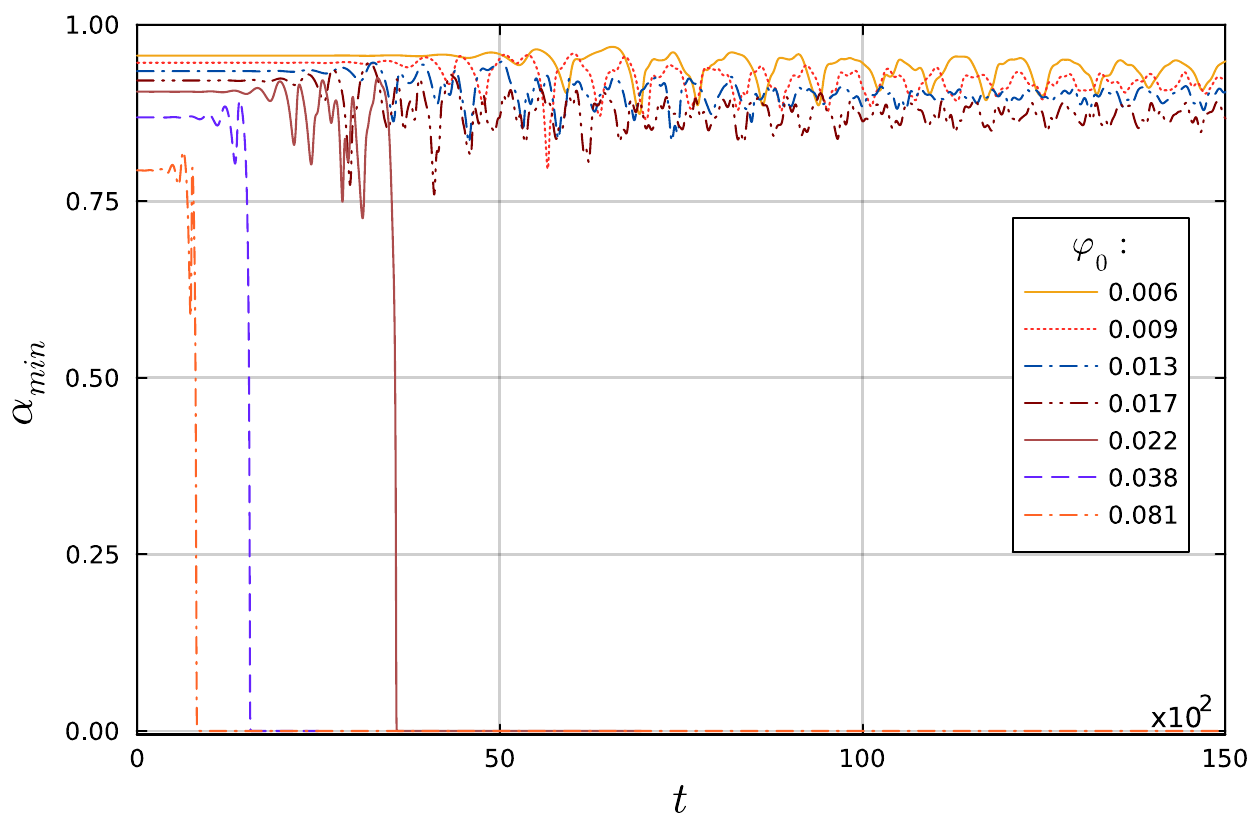}
\caption{Evolution of the central value of the lapse function $\alpha$ for the first excited state for all 7 models studied. For the 4 models with lower values of $\varphi_0$ migration is observed, while for models with higher values of $\varphi_0$ the central lapse collapses to zero indicating the formation of a black hole.}
\label{FirstState_Alpha_min}
\end{figure}

Figure~\ref{FirstState_Alpha_min} shows the value of the lapse function $\alpha$ at $r=0$ as a function of time for all 7 models during their evolution. The behavior of the lapse provides us with important information about the type of phenomena occurring during the evolution of our perturbed Proca stars. We observe two types of behavior corresponding to migration to a more stable solution, and collapse to a black hole. This does not imply that dispersion phenomena can not occur; in fact, we find that dispersion seems to predominate when the Proca stars have positive binding energy, but here we want to concentrate on the negative binding energy configurations.

\begin{figure}[!htbp]
\centering
\includegraphics[width=0.6\linewidth,height=0.7\textwidth]{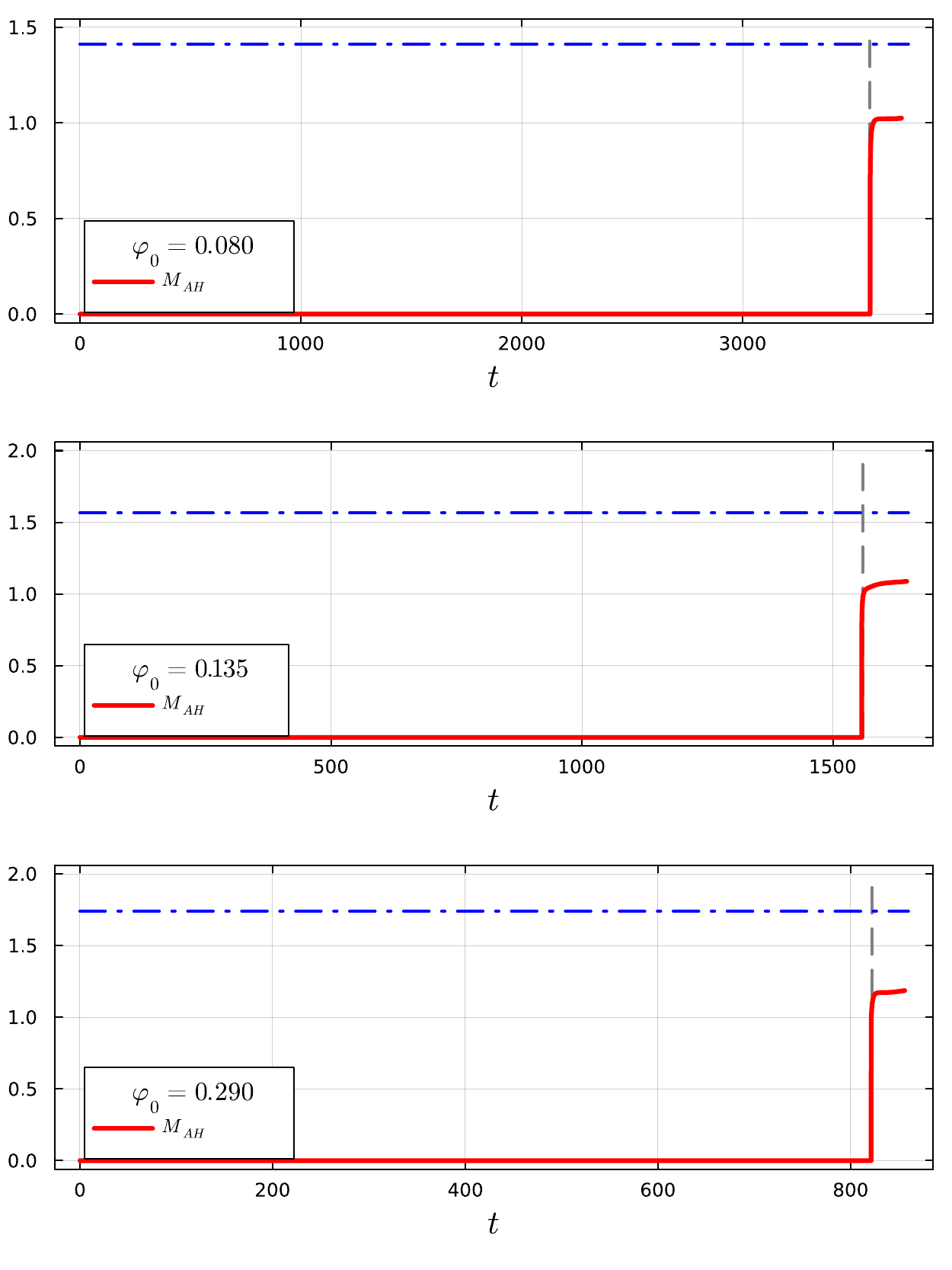}
\caption{Apparent horizon mass as a function of time time for the three models in the first excited state that exhibit collapse to a black hole. The time at which the code first finds an apparent horizon corresponds to the dashed vertical line. The dash-dot horizontal line represents the value of the total integrated mass of the spacetime at $t = 0$.}
\label{FirstState_AH}
\end{figure}

The central lapse collapses to zero for the models with $\varphi_0 =$ 0.081, 0.038, and 0.022, indicating the formation of a black hole. The behavior of the lapse is similar in all 3 models, the main difference being the time at which the collapse occurs. In particular, we find that if the star's mass is closer to the maximum (binding energy closer to the minimum) it becomes more unstable, leading to a faster collapse. In order to confirm the formation of a black hole we look for apparent horizons during the evolution. Figure~\ref{FirstState_AH} shows the apparent horizon mass as a function of time. We can observe that the apparent horizon suddenly appears in all three cases (at different times), coinciding with the time at which the central lapse collapses to zero. The horizon mass is always smaller than the initial total mass of the spacetime (shown as a blue dashed line). We only show the horizon mass for a short time after its formation, as later the numerical errors become too large due to the well known phenomenon of slice stretching (we are evolving without a shift vector).

\begin{figure}[!htbp]
\centering
\includegraphics[width=0.6\textwidth]{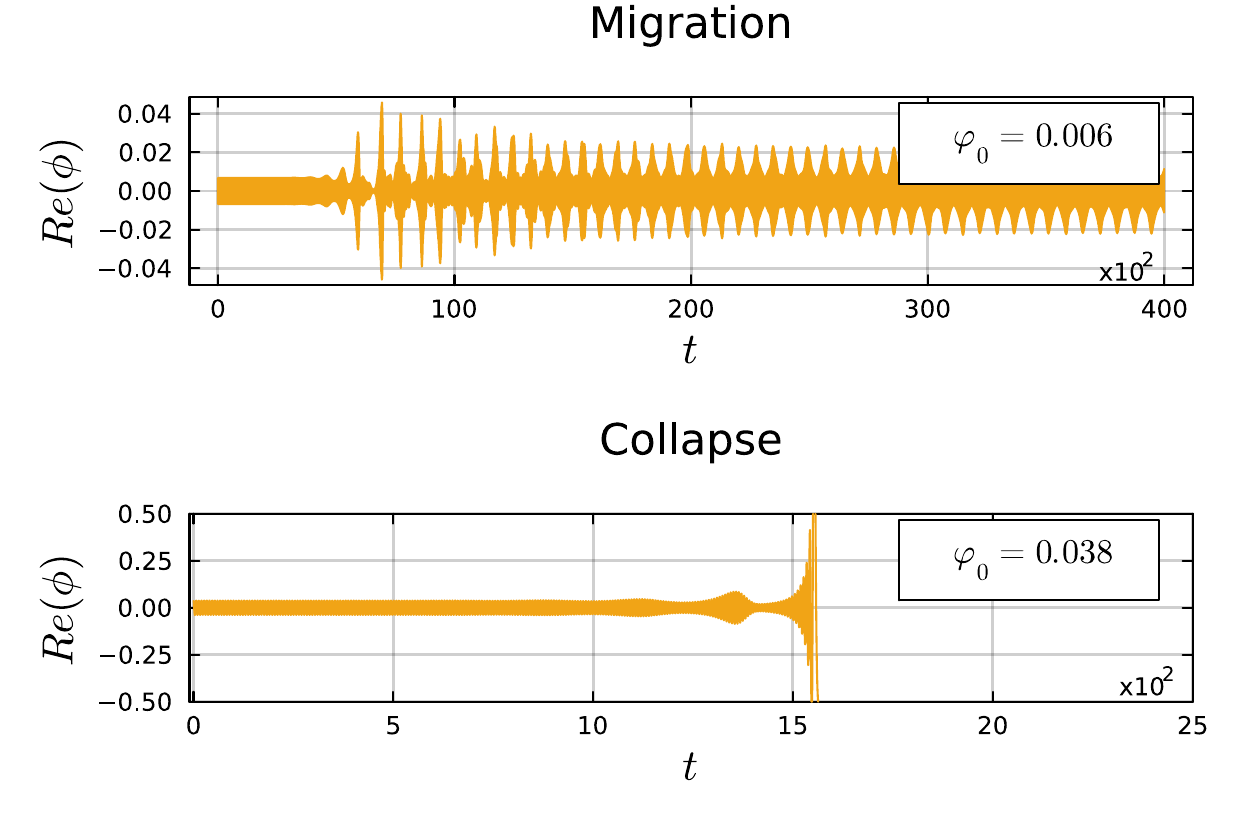}
\caption{Evolution of the real part of the scalar potential $\phi$ at $r=0$ for the models with $\varphi_0 = 0.006$ (upper panel) and $\varphi_0 = 0.038$ (lower panel). These models illustrate the behavior of the field's oscillations when the Proca star is either migrating to another more stable solution, or collapsing into a black hole.}
\label{FirstState_Phi_time}
\end{figure}

We will now focus on the migration phenomena which occurs for the models with $\varphi_0 =$ 0.017, 0.013, 0.009 and 0.006. These models have lower masses and a binding energy closer to zero. In all these cases the central value of the lapse begins to oscillate irregularly after some time around a mean value that stays away from zero, and is also lower than its original value.  However, this provides little information as to the final state of the Proca star. To gain a better understanding of what is happening, in Figure~\ref{FirstState_Phi_time} we show the evolution of the oscillations of the real part of the scalar potential $\phi$ evaluated at $r=0$ for the model with $\varphi_0 = 0.006$, which migrates (upper panel), and the model with $\varphi_0 = 0.038$, which collapses to a black hole (lower panel).  Notice that for the collapsing model the oscillations become very large and eventually stop due to the collapse of the lapse there, while for the migrating model the oscillations settle on a quite regular modulated pattern after some time. We observe a similar behavior for the other migrating models.

\begin{figure}[!htbp]
\centering
\includegraphics[width=0.6\textwidth]{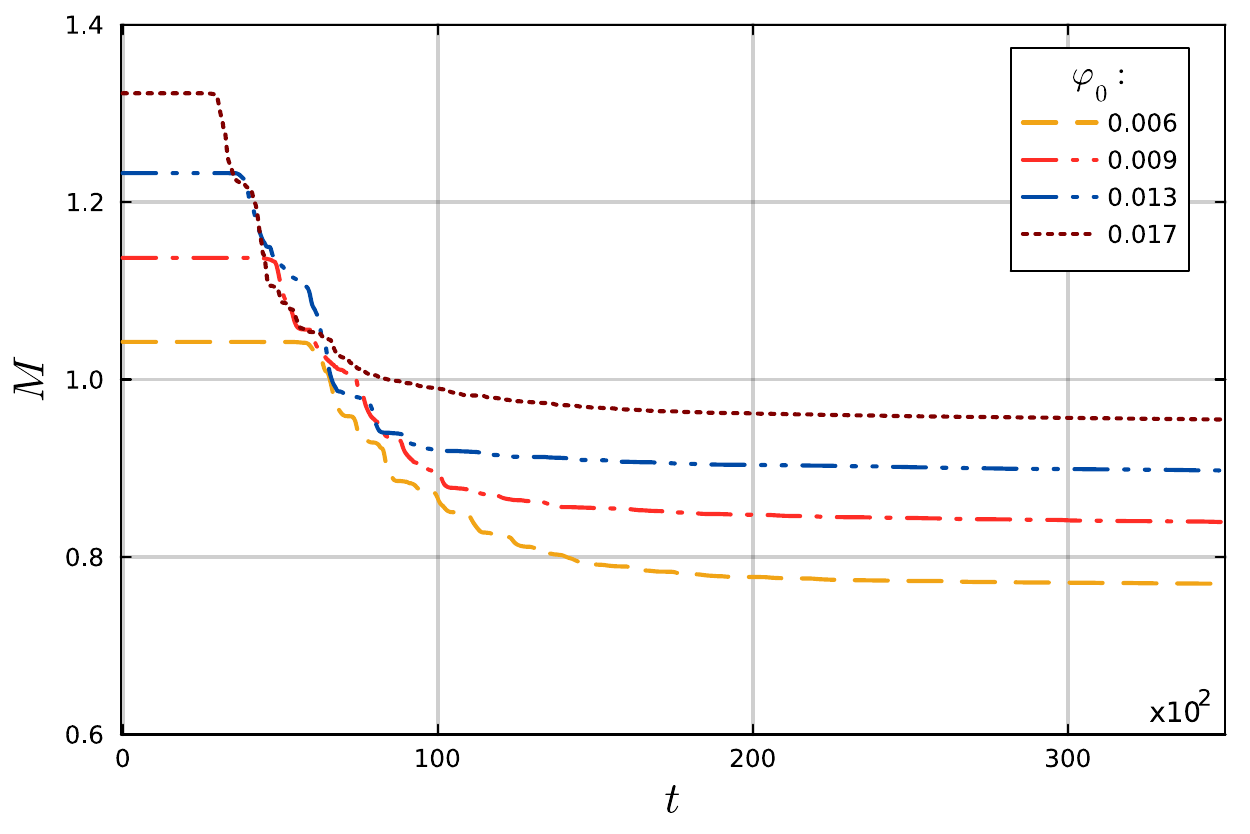}
\caption{Evolution of the total mass $M$ at the outer boundary for models in the first excited state that exhibit migration.}
\label{FirstState_MassBounder_Models}
\end{figure}

For the migrating models it is important to determine the direction of this migration, or in other words the final asymptotic state. In order to do this we analyze the total integrated mass at the boundary of our numerical domain for the 4 migrating models; this is shown in Figure~\ref{FirstState_MassBounder_Models}. The different configurations start with an initial total mass $M_{i}$ corresponding to that of the (slightly perturbed) stationary solution.  However, due to the perturbation, a fraction of this mass --- which in all 4 cases turns out to be approximately 25\% --- is lost during the evolution as the star ejects part of the Proca field to infinity. As a result, the total mass at the boundary decreases over time to some asymptotic value $M_{f}$.  Notice also that the total mass remains initially constant for a while, which corresponds to the time it takes for the ejected field to reach the boundary of our computational domain.

\begin{figure}[!htbp]
\centering
\includegraphics[width=0.6\textwidth]{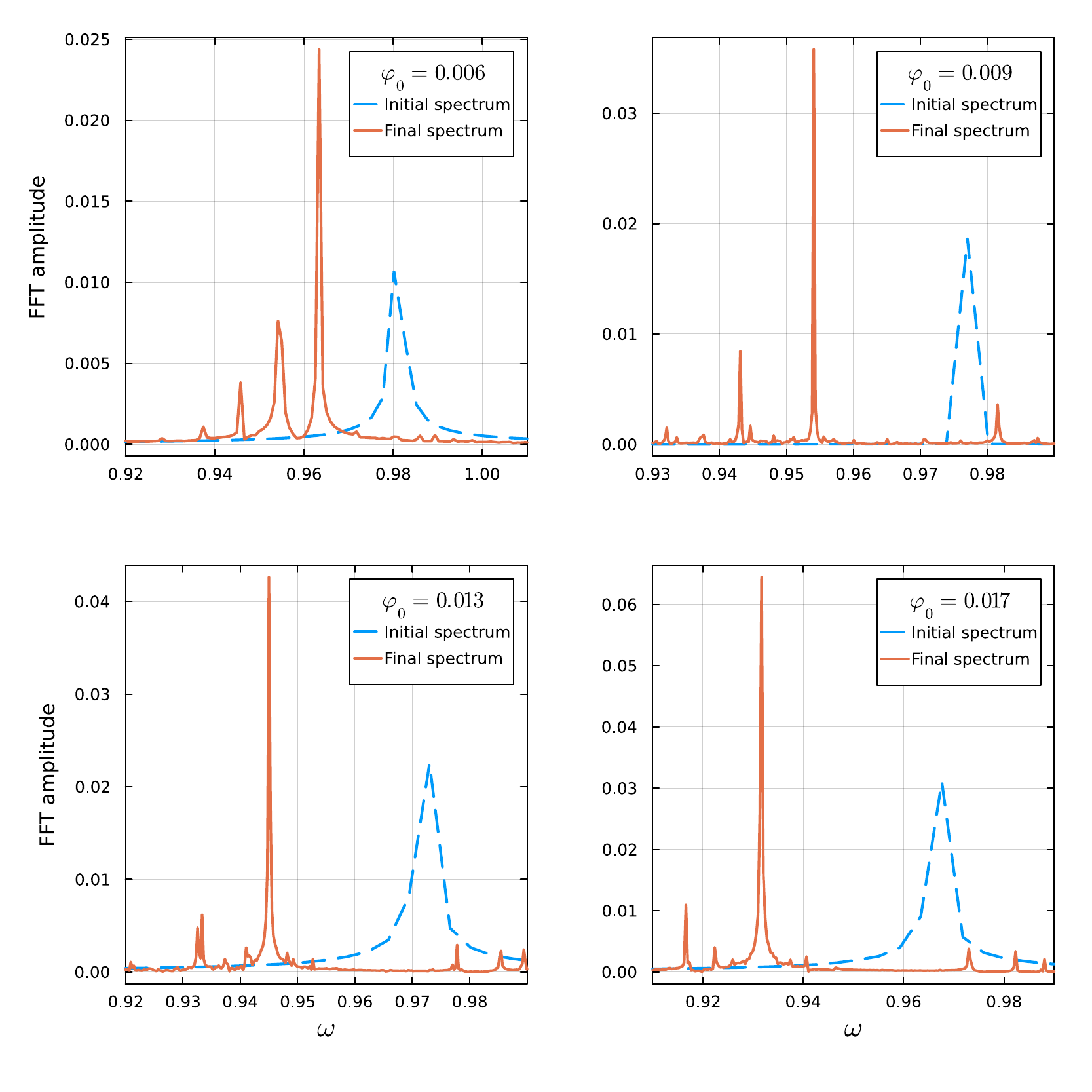}
\caption{Fourier transform of the time evolution of the real part of $\phi$ for models in the first excited state, specifically for those models that exhibit the migration phenomena. The dashed blue lines correspond to the initial frequency spectrum, while the solid red lines represent the final spectrum. In the final stage, the migrating stars have more than one frequency present, but there is always one dominant frequency that stands out.}
\label{FirstState_Subplot_FFT}
\end{figure}

We can analyze the final state of our configurations further by doing a Fast Fourier Transform (FFT) of the real part of the scalar potential $\phi$ evaluated at $r=0$, both at the beginning and at the final stages of the time evolutions.  This is shown in Figure~\ref{FirstState_Subplot_FFT}. The dashed blue line shows the FFT for the initial stages of the evolution for all 4 migrating models. In each case, we can see a single peak frequency $\omega_{i}$, which coincides with the oscillating frequency of our original stationary solution. The solid red lines correspond to the FFT for the late stages of our evolutions. The spectra for the different models have clearly changed in time, and now show a very different dominant frequency $\omega_{f}$, plus a few other lower amplitude modulating frequencies (we have checked that when doing longer simulations the height of those secondary peaks becomes smaller, showing that those modulating oscillations decay away slowly). To obtain the FFT's, we divide the evolution into initial and final stages based on the behavior of the oscillation of the real part of the scalar potential at $r=0$.  We select an appropriate time interval for each migrating case. Generally, the initial stage consists of only a short time interval because the oscillations of the perturbed solution increase rapidly. The final stage corresponds to a later time when the oscillations have settled into a regular modulated pattern. We only exclude a short portion of the evolution, which represents the transition phase in the middle of the evolutions.

\begin{figure}[!htbp]
\centering
\includegraphics[width=0.5\linewidth]{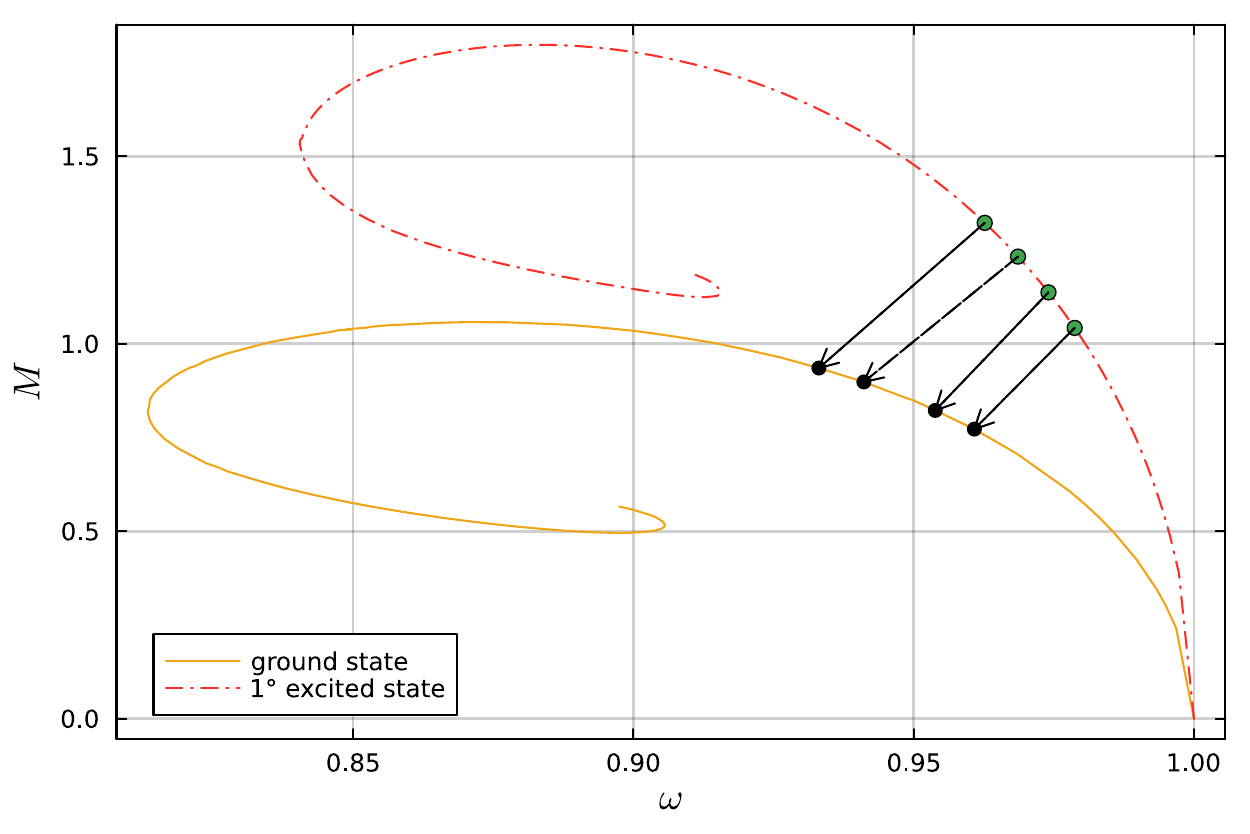}
\caption{Using the data for the final dominant frequency $\omega_{f}$ and final mass and $M_{f}$ for our migrating models in the first excited state (see Table~\ref{Table-Migration} of the Appendix), we can find the final configuration in the ground state family of solutions to which they are migrating.}
\label{Migration_gnd_exc1}
\end{figure}

By considering the final total mass $M_{f}$ and the final dominant frequency $\omega_{f}$, we can determine the direction of migration for the perturbed Proca stars in the first excited state. Upon comparing these final parameters with the families of stationary solutions, we conclude that the migration is towards configurations in the stable branch of the ground state, see Figure~\ref{Migration_gnd_exc1}. All this data is summarized in Table~\ref{Table-Migration} of the appendix.


\subsubsection{Second excited state}

\begin{figure}[!htbp]
\centering
\includegraphics[width=0.6\textwidth]{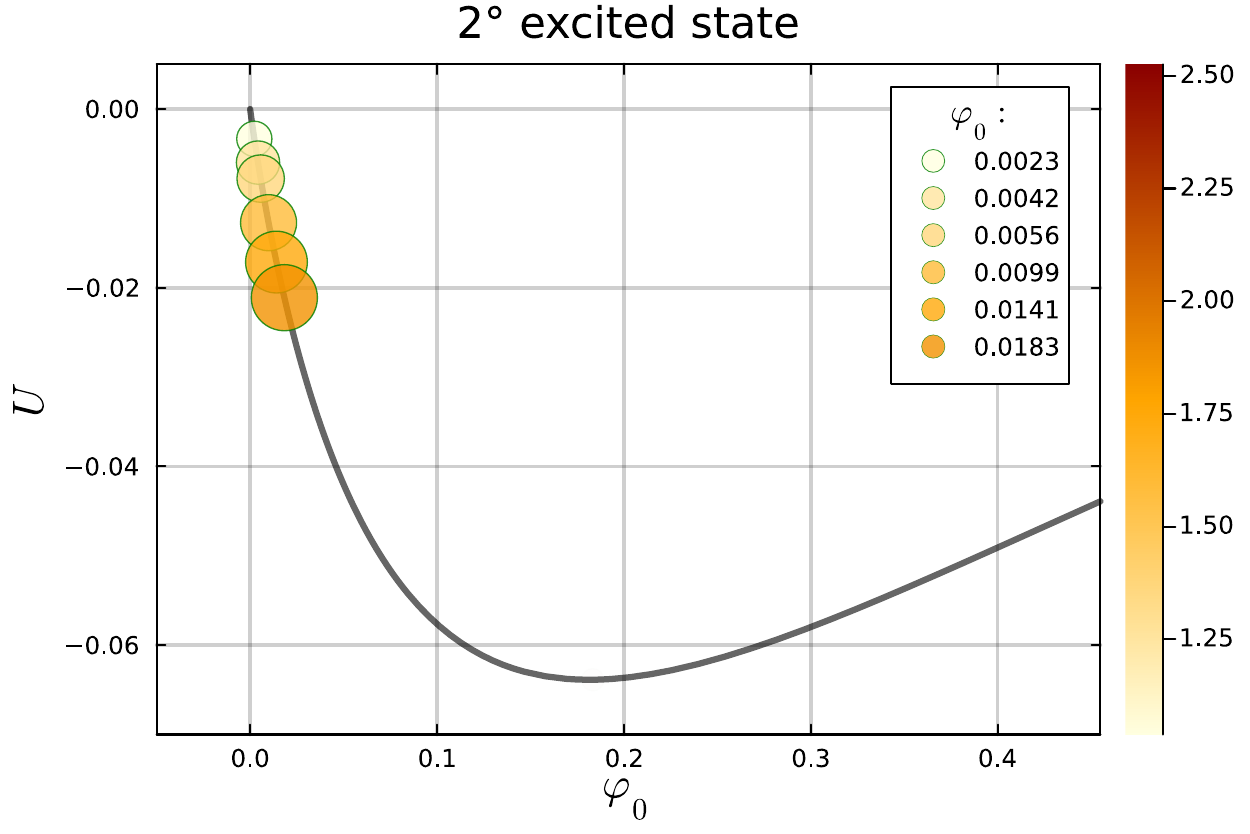}
\caption{Binding energy $U$ as a function of $\varphi_0$ for the second excited state. The circles show the 6 models that we considered for our perturbed dynamical evolutions. As before, the color bar and the size of the circles indicate the initial total mass.}
\label{SecondState_BE_Models}
\end{figure}

As shown in Figure~\ref{SecondState_BE_Models}, for the case of the second excited states we have also selected models with negative binding energy to the left of the minimum, but in this case even closer to the origin. The reason for this is the general trend of instability for the Proca stars: for higher excited states collapse to a black hole becomes more likely, and it is harder to find models that migrate to a stable configurations. 

\begin{figure}[!htbp]
\centering
\includegraphics[width=0.6\textwidth]{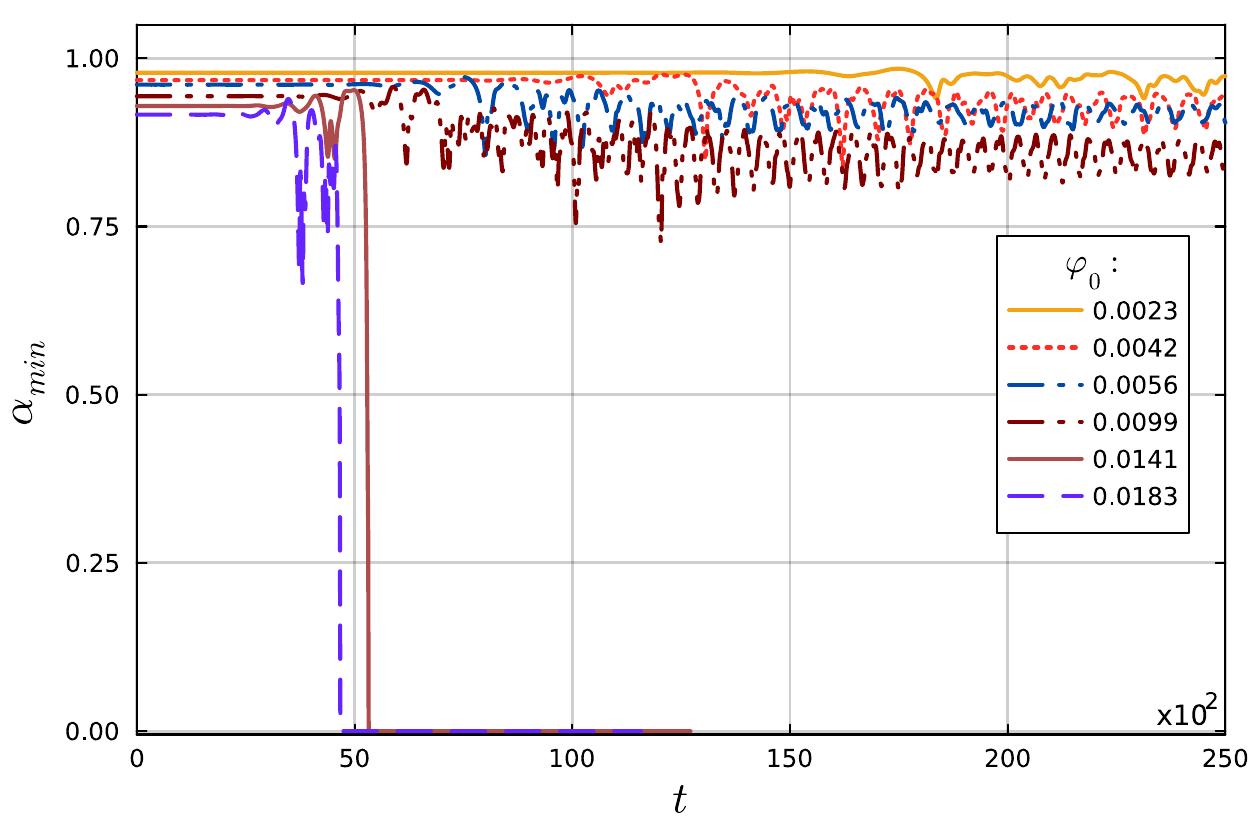}
\caption{Evolution of the central value of the lapse function $\alpha$ for the second excited state for all the 6 models studied. For the models with lower values of $\varphi_0$ migration is observed, while the remaining 2 models collapse into black hole.}
\label{SecondState_Alpha_min}
\end{figure}

\begin{figure}[!htbp]
\centering
\includegraphics[width=0.6\linewidth,height=0.7\textwidth]{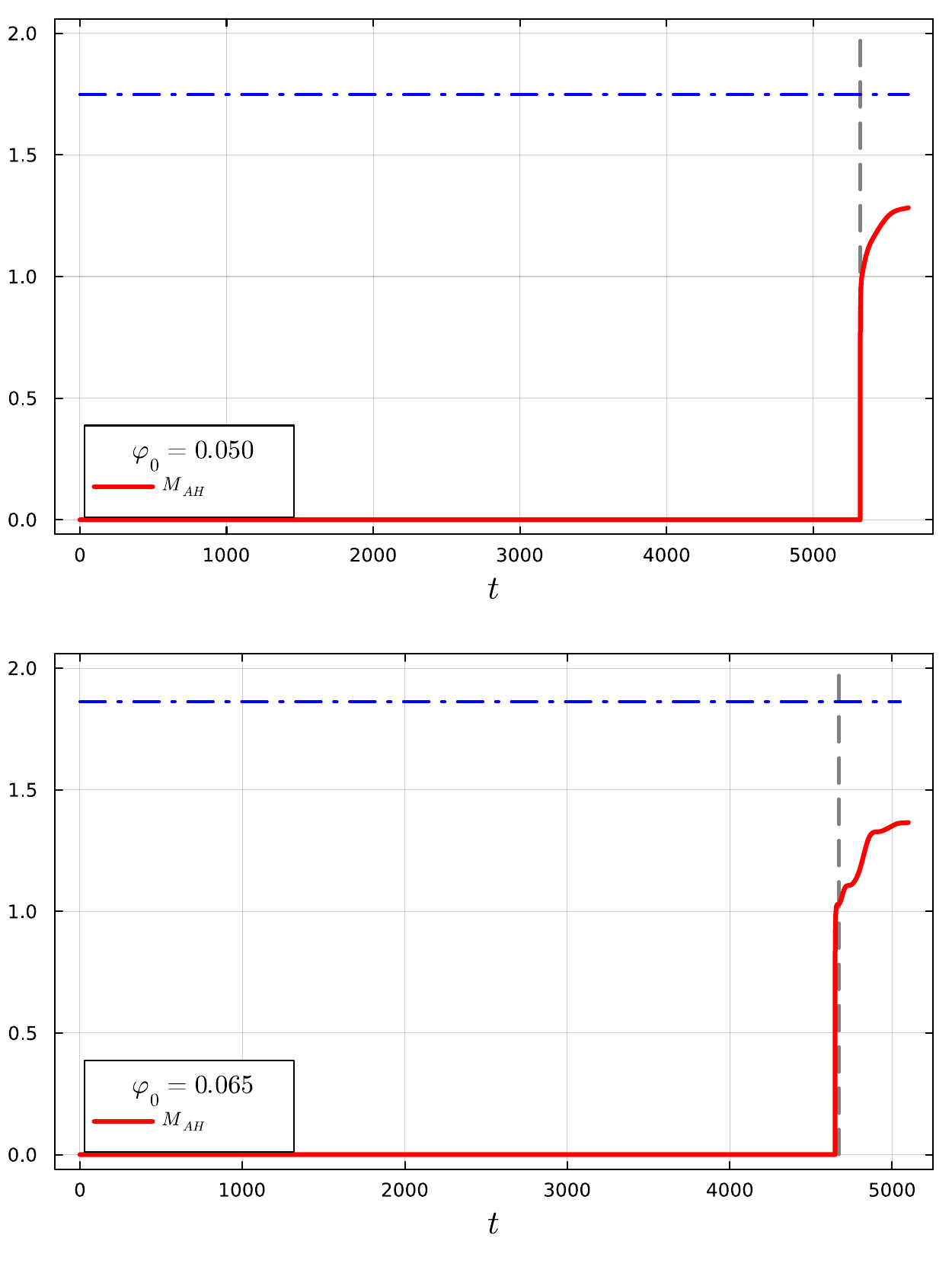}
\caption{Apparent horizon mass as a function of time for the 2 collapsing models in the second excited state. The dashed vertical line indicates the time at which an apparent horizon is first found, and the dashed-dot horizontal line represents the value of the total integrated mass at $t=0$.}
\label{SecondState_AH}
\end{figure}

Figure~\ref{SecondState_Alpha_min} shows again the central value of the lapse function $\alpha$ as a function of time for these 6 models. Similar behaviors are observed, with migration appearing for the 4 lower mass models, and collapse to a black hole for the other 2. Figure~\ref{SecondState_AH} shows the apparent horizon mass as a function of time for the two collapsing models. We can observe again that the apparent horizon suddenly appears at a time coinciding with the moment at which the central lapse collapses to zero. It is, however, important to mention that in fact the majority of Proca star configurations in the second excited state experience collapse to a black hole, with only configurations with very low masses potentially presenting signs of migration. The fact that 4 out of our 6 models migrate is because we have selected them on purpose.

\begin{figure}[!htbp]
\centering
\includegraphics[width=0.6\textwidth]{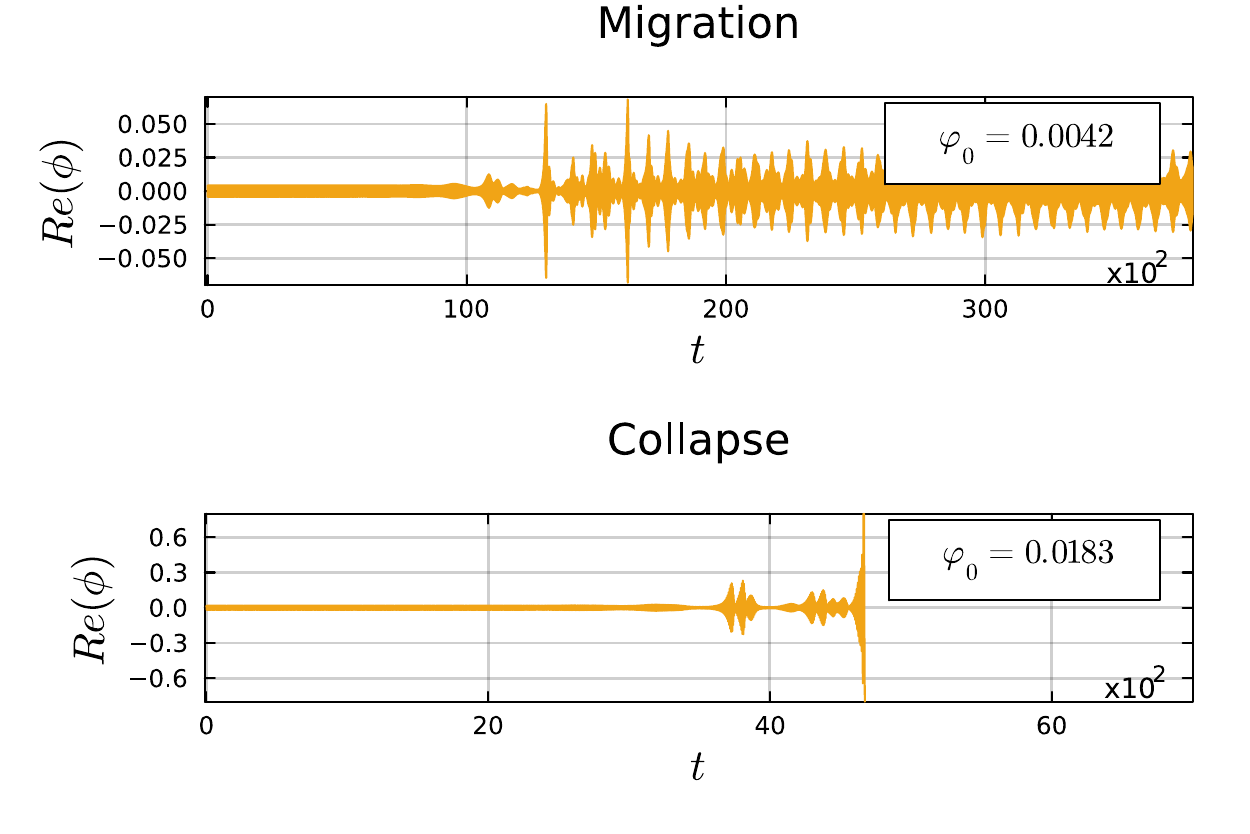}
\caption{Evolution of the real part of the scalar potential $\phi$ at $r=0$ in for the models with with $\varphi_0=0.0042$ (upper panel) and $\varphi_0$ = 0.0183 (lower panel).
These models illustrate the behavior of the field's oscillations when the PSs are either migrating to a more stable solution or collapsing to a black hole.}
\label{SecondState_Phi_time}
\end{figure}

Figure~\ref{SecondState_Phi_time} shows the oscillations of the real part of the scalar potential $\phi$ evaluated at $r=0$ for the models with $\varphi_0=0.0042$ and $\varphi_0$ = 0.0183.  Again we see how for the collapsing model (lower panel) the oscillations increase in amplitude and later stop due to the collapse of the lapse, while for the migrating model (upper panel) the oscillations at late times settle on a simple modulated oscillation.

\begin{figure}[!htbp]
\centering
\includegraphics[width=0.6\textwidth]{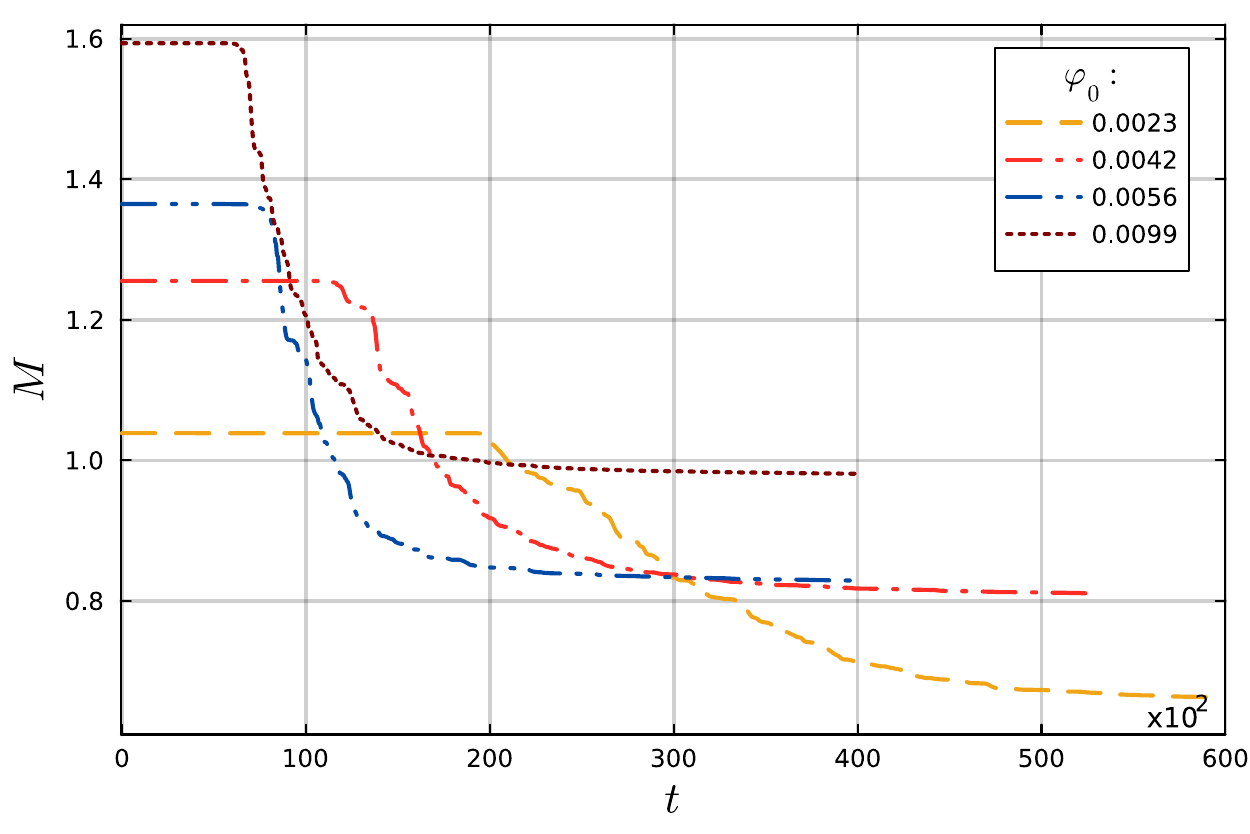}
\caption{Evolution of the total mass $M$ at the boundary for models in the second excited state that exhibit migration.}
\label{SecondState_MassBounder_Models}
\end{figure}

Let us now return to the 4 migrating models with $\varphi_0 =$ 0.0099, 0.0056, 0.0042 and 0.0023. All of these models display irregular oscillations in the central value of the lapse function around some equilibrium position. The total integrated mass at the boundary as a function of time for these models is shown in Figure~\ref{SecondState_MassBounder_Models}. Here, we again observe a fractional loss of the initial mass due to the fact that some of the field is ejected to infinity (in all these cases approximately 35-39\%).


\begin{figure}[!htbp]
\centering
\includegraphics[width=0.6\textwidth]{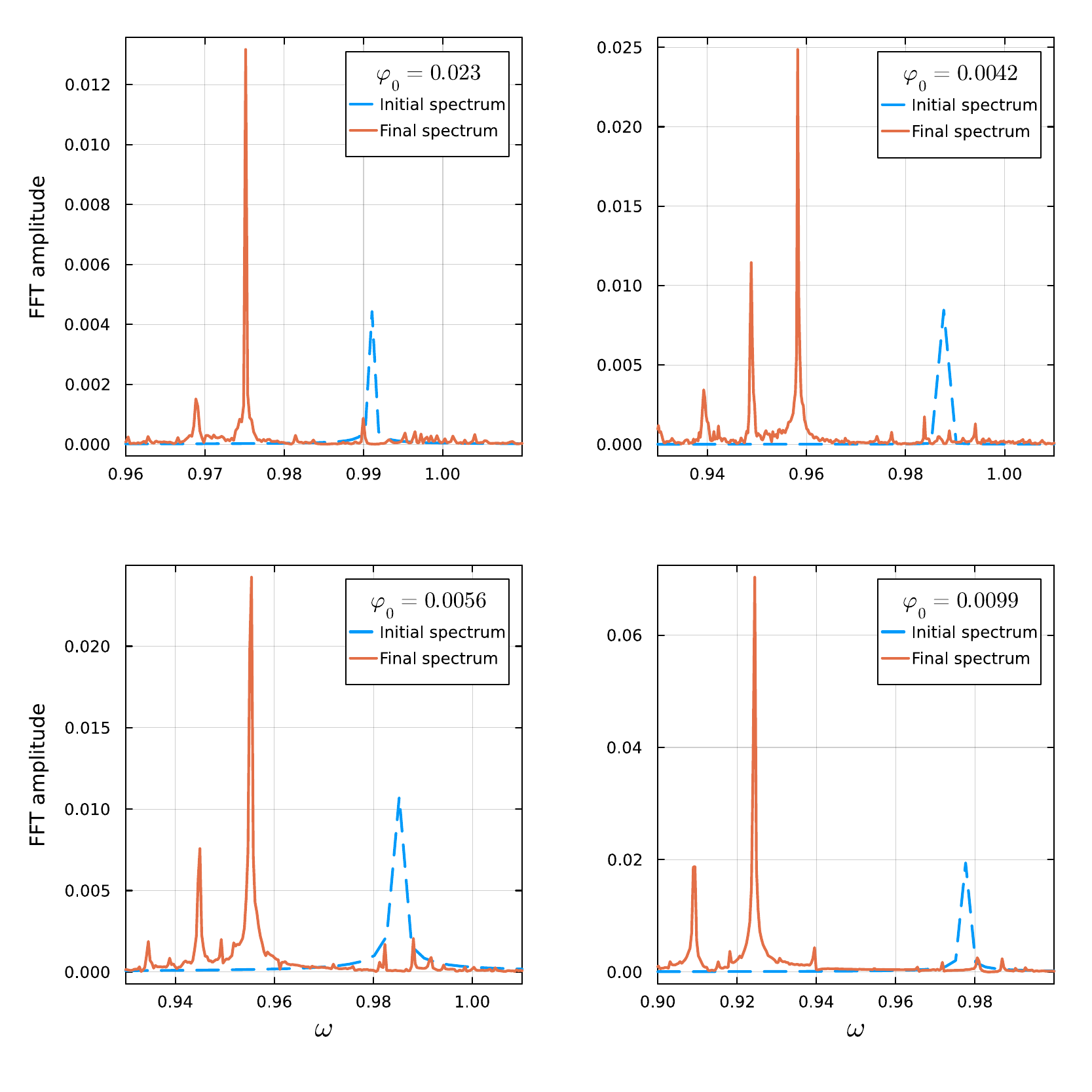}
\caption{Fourier transform of the time evolution of the real part of $\phi$ evaluated at $r=0$ for those models in the second excited state that exhibit the migration phenomena. These results are similar to those for the first excited state.}
\label{SecondState_Subplot_FFT}
\end{figure}

\begin{figure}[!htbp]
\centering
\includegraphics[width=0.6\linewidth]{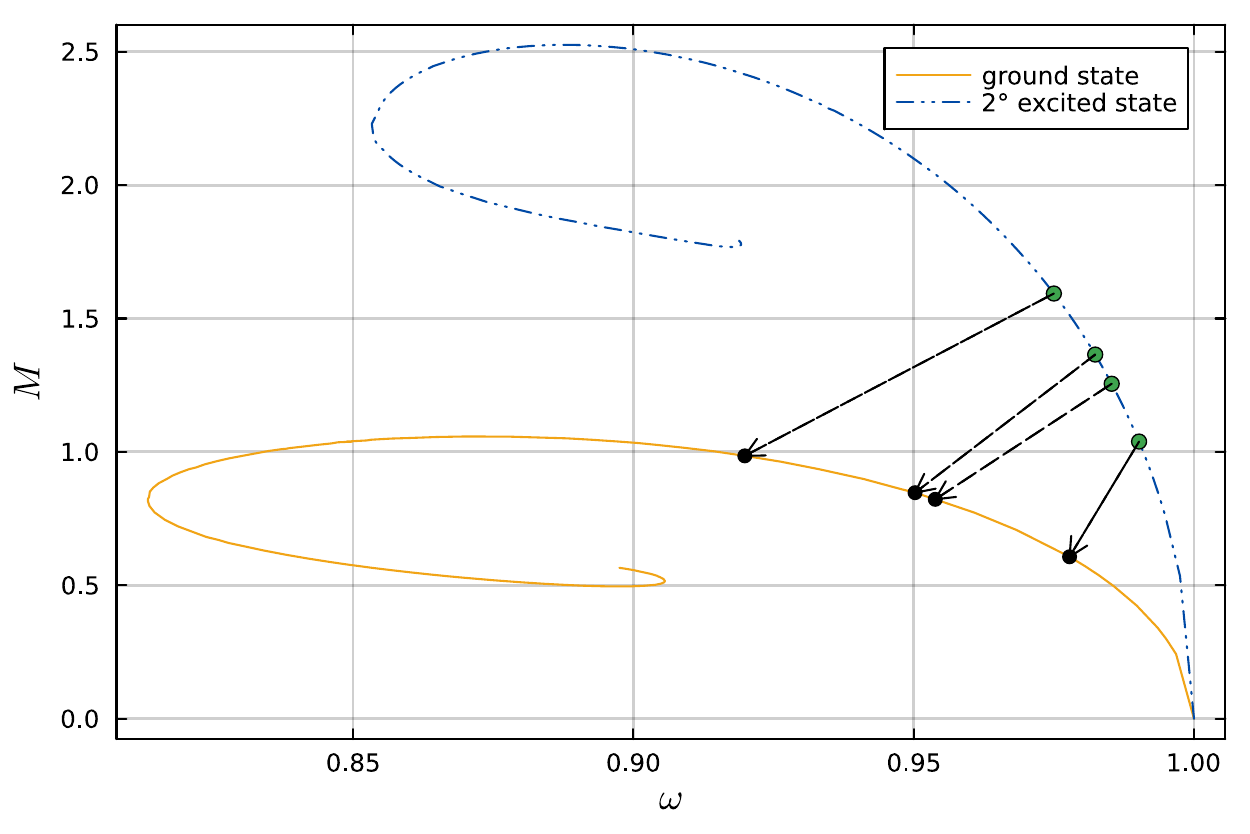}
\caption{Using the data for the final dominant frequency $\omega_{f}$ and final mass and $M_{f}$  for our migrating models in the second excited state (shown in Table~\ref{Table-Migration} of the Appendix), we can find the final configuration in the ground state family of solutions to which they are migrating.}
\label{Migration_gnd_exc2}
\end{figure}

We also obtain the FFT of the real part of $\phi$ evaluated at $r=0$ for the initial and final stages of the evolution, see Figure~\ref{SecondState_Subplot_FFT}. The FFT analysis reveals the frequency spectra from which we can obtain the dominant initial and final frequencies $\omega_{i}$ and $\omega_{f}$. As before, by comparing the total final mass $M_{f}$ and the final dominant frequency $\omega_{f}$ with the families of stationary solutions, we can conclude that the stars are migrating to stable configurations on the ground state, see Figure~\ref{Migration_gnd_exc2}. All the pertinent data for the second excited state is summarized in Table~\ref{Table-Migration} of the appendix.

\section{Discussion and conclusions}
\label{sec05:conclusions}

In this paper we have considered families of solutions for excited states of Proca stars in spherical symmetry, with a focus on exploring the dynamical behavior of the first two excited states under perturbations. We considered both the families of stationary solutions, as well as non-linear dynamical simulations of a set of self-consistent perturbed models. Our findings offer valuable insights into the stability and evolution of excited Proca stars, contributing to a further understanding of these compact objects.

It is well known that spherical Proca stars in the ground state are stable against small perturbations for a specific branch corresponding to those configurations with negative binding energy and mass to the left of the maximum mass (minimum binding energy) in the mass versus $\varphi_0$ diagram.  However, we find that in the case of Proca stars in the first and second excited states this stable branch does not exist.  In fact, we find that for excited states only a \textit{metastable} region exists for configurations with negative binding energy and low masses. In this metastable branch the Proca stars migrate under perturbations to solutions with lower mass in the stable branch of the ground state, while ejecting excess field to infinity. The behavior of excited Proca stars is then very similar to the one found in the case of scalar boson stars, where dynamical evolutions have shown that excited states are unstable, and they either decay to the fundamental state or collapse to a black hole~\cite{Balakrishna:1997ej}. Our dynamical simulations of excited Proca stars have allowed us not only to observe the migration phenomena but also to determine the final configuration on the ground state to which they are migrating. 

We should stress the fact that our models do not incorporate self-interactions. It has been shown that self-interactions can stabilize excited states of scalar boson stars~\cite{Sanchis_Gual_2022,DiGiovanni:2021vlu}.  However, recent studies demonstrate that in the case of vector fields self-interactions can lead to significant pathologies, particularly the loss of hyperbolicity of the field equations~\cite{Clough_2022,Coates_2022}. Introducing self-interaction parameters into the Proca star model therefore requires careful consideration of these instabilities, as highlighted by recent findings~\cite{Aoki:2022woy,Herdeiro:2023a,Brito:2024,Rubio:2024ryv}. This presents an opportunity to further explore the stability and dynamics of excited Proca stars, and to understand the complexities associated with self-interactions. We are currently in the process of studying this further.


\acknowledgments

We thank Jose Damian Lopez, Claudio Lazarte, Axel Rangel and Jorge Yahir Mio for many useful discussions and comments. This work was partially supported by CONAHCYT Network Projects No. 376127 and No. 304001, and DGAPA-UNAM project IN100523.


\appendix


\section{Migration data}

In Table~\ref{Table-Migration} we present the data corresponding to the models in the first and second excited states that exhibit migration phenomena. We first show the initial values of the frequency $\omega_i$ and total mass $M_i$ at $t=0$, representing the parameters of the unperturbed stationary configuration. Subsequently, we show the final parameters $\omega_f$ and $M_f$ at the late stage of our dynamical evolutions, as well as their respective uncertainties due to numerical error.  We estimate these uncertainties in $\omega_f$ from the width of the peak in the FFT, and in $M_f$ from the slow decay of the total mass at the end of our simulations toward an asymptotic value. In the latter case, the uncertainty of $\pm 0.005$ corresponds to the last decimal place for which the value of the mass is still changing, {\em i.e.} in all our simulations the third decimal place in the value of $M_f$ is still decreasing very slowly at the end, so the asymptotic value should be closer to the lower bound. A comparison of the initial and final parameters reveals notable changes in both the frequency of oscillation of the Proca field and the total integrated mass. 

The last two columns in the table show the frequency $\omega$ and mass $M$ for those solutions in the ground state toward which the perturbed excited solutions seem to be migrating. It is important to mention the fact that the frequencies of the ground state solutions $\omega$ used for the comparisons are not exactly the same as those of the final stage of the migrating solutions $\omega_f$.  The reason for this is that we do not have solutions for all values of $\omega$ since we have only sampled the solution space for a finite (though large) number of cases, and the frequency is not our free parameter, but rather the eigenvalue found at the end. However, we choose those solutions in our family of sampled ground state configurations for which the frequency is closest to $\omega_f$.

With these caveats, once we choose the comparison frequency $\omega$ for a ground state solution, the corresponding mass $M$ is fixed, and it is remarkable how close it is to the value $M_f$ of the migrating solution in each case. This convinces us that the perturbed excited states are indeed migrating to stable solutions on the ground state.

\begin{table}
\centering
\begin{tabular}{|>{\centering\arraybackslash}m{6em}||>{\centering\arraybackslash}m{1.5cm}|>{\centering\arraybackslash}m{1.5cm}|>{\centering\arraybackslash}m{3cm}|>{\centering\arraybackslash}m{2cm}||>{\centering\arraybackslash}m{1.5cm}|>{\centering\arraybackslash}m{1.5cm}|}
\hline 
\textbf{Models} & \textbf{$\omega_{i}$} & \textbf{$M_{i}$} & \textbf{$\omega_{f} \pm \delta \omega_{f}$} & $M_{f} \pm$ 0.005 & \textbf{$\omega$} & \textbf{$M$} \\ 
\hline\hline
1° excited & \multicolumn{4}{c||}{Migration} & \multicolumn{2}{c|}{Initial data: gnd} \\ 
\hline 
0.006 & 0.980 & 1.042 & 0.96287 $\pm$  0.00031 & 0.769 & 0.960 & 0.772\\ 
\hline 
0.009 & 0.977 & 1.137 & 0.95407 $\pm$ 0.00024 & 0.839 & 0.953 & 0.822\\ 
\hline 
0.013 & 0.973 & 1.232 & 0.94499 $\pm$ 0.00026 & 0.897 & 0.941 & 0.898\\ 
\hline 
0.017 & 0.967 & 1.322 & 0.93170 $\pm$ 0.00035 & 0.955 & 0.933 & 0.935\\ 
\hline\hline 
2° excited & \multicolumn{4}{c||}{Migration} & \multicolumn{2}{c|}{Initial data: gnd}\\ 
\hline
0.0023 & 0.990 & 1.038 & 0.97513 $\pm$ 0.00025 & 0.655 & 0.977 & 0.607\\ 
\hline 
0.0042 & 0.987 & 1.255 & 0.95823 $\pm$ 0.00022 & 0.811 & 0.953 & 0.822\\ 
\hline 
0.0056 & 0.985 & 1.364 & 0.95538 $\pm$ 0.00035 & 0.829 & 0.950 & 0.847\\ 
\hline 
0.0099 & 0.977 & 1.593 & 0.92446 $\pm$ 0.00041 & 0.981 & 0.919 & 0.985\\ 
\hline 
\end{tabular} 
\caption{Migration phenomena: For the first and second excited states, we show in the first two columns the initial frequency $\omega_i$ and initial total mass $M_i$ of the corresponding unperturbed configuration.  On the third and fourth columns we show the final dominant frequency $\omega_f$ and final mass $M_f$ after the migration has occurred.  The last two columns show the corresponding values for the frequency and mass of configurations on the stable branch of the ground state toward which the dynamical solutions seem to migrate.}
\label{Table-Migration}
\end{table}







\bibliographystyle{apsrev4-1}
\bibliography{referencias}

\begin{thebibliography}{53}%
\makeatletter
\providecommand \@ifxundefined [1]{%
 \@ifx{#1\undefined}
}%
\providecommand \@ifnum [1]{%
 \ifnum #1\expandafter \@firstoftwo
 \else \expandafter \@secondoftwo
 \fi
}%
\providecommand \@ifx [1]{%
 \ifx #1\expandafter \@firstoftwo
 \else \expandafter \@secondoftwo
 \fi
}%
\providecommand \natexlab [1]{#1}%
\providecommand \enquote  [1]{``#1''}%
\providecommand \bibnamefont  [1]{#1}%
\providecommand \bibfnamefont [1]{#1}%
\providecommand \citenamefont [1]{#1}%
\providecommand \href@noop [0]{\@secondoftwo}%
\providecommand \href [0]{\begingroup \@sanitize@url \@href}%
\providecommand \@href[1]{\@@startlink{#1}\@@href}%
\providecommand \@@href[1]{\endgroup#1\@@endlink}%
\providecommand \@sanitize@url [0]{\catcode `\\12\catcode `\$12\catcode
  `\&12\catcode `\#12\catcode `\^12\catcode `\_12\catcode `\%12\relax}%
\providecommand \@@startlink[1]{}%
\providecommand \@@endlink[0]{}%
\providecommand \url  [0]{\begingroup\@sanitize@url \@url }%
\providecommand \@url [1]{\endgroup\@href {#1}{\urlprefix }}%
\providecommand \urlprefix  [0]{URL }%
\providecommand \Eprint [0]{\href }%
\providecommand \doibase [0]{http://dx.doi.org/}%
\providecommand \selectlanguage [0]{\@gobble}%
\providecommand \bibinfo  [0]{\@secondoftwo}%
\providecommand \bibfield  [0]{\@secondoftwo}%
\providecommand \translation [1]{[#1]}%
\providecommand \BibitemOpen [0]{}%
\providecommand \bibitemStop [0]{}%
\providecommand \bibitemNoStop [0]{.\EOS\space}%
\providecommand \EOS [0]{\spacefactor3000\relax}%
\providecommand \BibitemShut  [1]{\csname bibitem#1\endcsname}%
\let\auto@bib@innerbib\@empty
\bibitem [{\citenamefont {Kaup}(1968)}]{Kaup68}%
  \BibitemOpen
  \bibfield  {author} {\bibinfo {author} {\bibfnamefont {D.~J.}\ \bibnamefont
  {Kaup}},\ }\href@noop {} {\bibfield  {journal} {\bibinfo  {journal} {Phys.
  Rev.}\ }\textbf {\bibinfo {volume} {172}},\ \bibinfo {pages} {1331} (\bibinfo
  {year} {1968})}\BibitemShut {NoStop}%
\bibitem [{\citenamefont {Ruffini}\ and\ \citenamefont
  {Bonazzola}(1969)}]{Ruffini69}%
  \BibitemOpen
  \bibfield  {author} {\bibinfo {author} {\bibfnamefont {R.}~\bibnamefont
  {Ruffini}}\ and\ \bibinfo {author} {\bibfnamefont {S.}~\bibnamefont
  {Bonazzola}},\ }\href@noop {} {\bibfield  {journal} {\bibinfo  {journal}
  {Phys. Rev.}\ }\textbf {\bibinfo {volume} {187}},\ \bibinfo {pages} {1767}
  (\bibinfo {year} {1969})}\BibitemShut {NoStop}%
\bibitem [{\citenamefont {Brito}\ \emph
  {et~al.}(2016{\natexlab{a}})\citenamefont {Brito}, \citenamefont {Cardoso},
  \citenamefont {Herdeiro},\ and\ \citenamefont {Radu}}]{BRITO2016291}%
  \BibitemOpen
  \bibfield  {author} {\bibinfo {author} {\bibfnamefont {R.}~\bibnamefont
  {Brito}}, \bibinfo {author} {\bibfnamefont {V.}~\bibnamefont {Cardoso}},
  \bibinfo {author} {\bibfnamefont {C.~A.}\ \bibnamefont {Herdeiro}}, \ and\
  \bibinfo {author} {\bibfnamefont {E.}~\bibnamefont {Radu}},\ }\href {\doibase
  https://doi.org/10.1016/j.physletb.2015.11.051} {\bibfield  {journal}
  {\bibinfo  {journal} {Physics Letters B}\ }\textbf {\bibinfo {volume}
  {752}},\ \bibinfo {pages} {291} (\bibinfo {year}
  {2016}{\natexlab{a}})}\BibitemShut {NoStop}%
\bibitem [{\citenamefont {Liebling}\ and\ \citenamefont
  {Palenzuela}(2023)}]{Liebling_2023}%
  \BibitemOpen
  \bibfield  {author} {\bibinfo {author} {\bibfnamefont {S.~L.}\ \bibnamefont
  {Liebling}}\ and\ \bibinfo {author} {\bibfnamefont {C.}~\bibnamefont
  {Palenzuela}},\ }\href {http://dx.doi.org/10.1007/s41114-023-00043-4}
  {\bibfield  {journal} {\bibinfo  {journal} {Living Reviews in Relativity}\
  }\textbf {\bibinfo {volume} {26}} (\bibinfo {year} {2023})}\BibitemShut
  {NoStop}%
\bibitem [{\citenamefont {Schunck}\ and\ \citenamefont
  {Mielke}(2003)}]{Schunck_2003}%
  \BibitemOpen
  \bibfield  {author} {\bibinfo {author} {\bibfnamefont {F.~E.}\ \bibnamefont
  {Schunck}}\ and\ \bibinfo {author} {\bibfnamefont {E.~W.}\ \bibnamefont
  {Mielke}},\ }\href {http://dx.doi.org/10.1088/0264-9381/20/20/201} {\bibfield
   {journal} {\bibinfo  {journal} {Classical and Quantum Gravity}\ }\textbf
  {\bibinfo {volume} {20}} (\bibinfo {year} {2003})}\BibitemShut {NoStop}%
\bibitem [{\citenamefont {Seidel}\ and\ \citenamefont {Suen}(1991)}]{Seidel91}%
  \BibitemOpen
  \bibfield  {author} {\bibinfo {author} {\bibfnamefont {E.}~\bibnamefont
  {Seidel}}\ and\ \bibinfo {author} {\bibfnamefont {W.-M.}\ \bibnamefont
  {Suen}},\ }\href {https://link.aps.org/doi/10.1103/PhysRevLett.66.1659}
  {\bibfield  {journal} {\bibinfo  {journal} {Phys. Rev. Lett.}\ }\textbf
  {\bibinfo {volume} {66}},\ \bibinfo {pages} {1659} (\bibinfo {year}
  {1991})}\BibitemShut {NoStop}%
\bibitem [{\citenamefont {Finster}\ \emph {et~al.}(1999)\citenamefont
  {Finster}, \citenamefont {Smoller},\ and\ \citenamefont
  {Yau}}]{Finster:1998ws}%
  \BibitemOpen
  \bibfield  {author} {\bibinfo {author} {\bibfnamefont {F.}~\bibnamefont
  {Finster}}, \bibinfo {author} {\bibfnamefont {J.}~\bibnamefont {Smoller}}, \
  and\ \bibinfo {author} {\bibfnamefont {S.-T.}\ \bibnamefont {Yau}},\ }\href
  {\doibase 10.1103/PhysRevD.59.104020} {\bibfield  {journal} {\bibinfo
  {journal} {Phys. Rev. D}\ }\textbf {\bibinfo {volume} {59}},\ \bibinfo
  {pages} {104020} (\bibinfo {year} {1999})},\ \Eprint
  {http://arxiv.org/abs/gr-qc/9801079} {arXiv:gr-qc/9801079} \BibitemShut
  {NoStop}%
\bibitem [{\citenamefont {Liebling}\ and\ \citenamefont
  {Palenzuela}(2012)}]{Liebling:2012fv}%
  \BibitemOpen
  \bibfield  {author} {\bibinfo {author} {\bibfnamefont {S.~L.}\ \bibnamefont
  {Liebling}}\ and\ \bibinfo {author} {\bibfnamefont {C.}~\bibnamefont
  {Palenzuela}},\ }\href@noop {} {\bibfield  {journal} {\bibinfo  {journal}
  {Living Rev.Rel.}\ }\textbf {\bibinfo {volume} {15}},\ \bibinfo {pages} {6}
  (\bibinfo {year} {2012})},\ \Eprint {http://arxiv.org/abs/1202.5809}
  {arXiv:1202.5809 [gr-qc]} \BibitemShut {NoStop}%
\bibitem [{\citenamefont {Herdeiro}\ and\ \citenamefont
  {Radu}(2020)}]{Herdeiro:2020jzx}%
  \BibitemOpen
  \bibfield  {author} {\bibinfo {author} {\bibfnamefont {C.~A.~R.}\
  \bibnamefont {Herdeiro}}\ and\ \bibinfo {author} {\bibfnamefont
  {E.}~\bibnamefont {Radu}},\ }\href {\doibase 10.3390/sym12122032} {\bibfield
  {journal} {\bibinfo  {journal} {Symmetry}\ }\textbf {\bibinfo {volume}
  {12}},\ \bibinfo {pages} {2032} (\bibinfo {year} {2020})},\ \Eprint
  {http://arxiv.org/abs/2012.03595} {arXiv:2012.03595 [gr-qc]} \BibitemShut
  {NoStop}%
\bibitem [{\citenamefont {Mourelle}\ and\ \citenamefont
  {Adam}(2024)}]{Mourelle_2024}%
  \BibitemOpen
  \bibfield  {author} {\bibinfo {author} {\bibfnamefont {J.~C.}\ \bibnamefont
  {Mourelle}}\ and\ \bibinfo {author} {\bibfnamefont {C.}~\bibnamefont
  {Adam}},\ }\href {https://arxiv.org/abs/2407.07839} {\enquote {\bibinfo
  {title} {Galactic halos and rotating bosonic dark matter},}\ } (\bibinfo
  {year} {2024}),\ \Eprint {http://arxiv.org/abs/2407.07839} {2407.07839
  [astro-ph.GA]} \BibitemShut {NoStop}%
\bibitem [{\citenamefont {Sengo}\ \emph {et~al.}(2024)\citenamefont {Sengo},
  \citenamefont {Cunha}, \citenamefont {Herdeiro},\ and\ \citenamefont
  {Radu}}]{Sengo_2024}%
  \BibitemOpen
  \bibfield  {author} {\bibinfo {author} {\bibfnamefont {I.}~\bibnamefont
  {Sengo}}, \bibinfo {author} {\bibfnamefont {P.~V.~P.}\ \bibnamefont {Cunha}},
  \bibinfo {author} {\bibfnamefont {C.~A.~R.}\ \bibnamefont {Herdeiro}}, \ and\
  \bibinfo {author} {\bibfnamefont {E.}~\bibnamefont {Radu}},\ }\href
  {https://arxiv.org/abs/2402.14919} {\enquote {\bibinfo {title} {The imitation
  game reloaded: effective shadows of dynamically robust spinning proca
  stars},}\ } (\bibinfo {year} {2024}),\ \Eprint
  {http://arxiv.org/abs/2402.14919} {2402.14919 [gr-qc]} \BibitemShut {NoStop}%
\bibitem [{\citenamefont {Rosa}\ and\ \citenamefont
  {Rubiera-Garcia}(2022)}]{Rosa_2022}%
  \BibitemOpen
  \bibfield  {author} {\bibinfo {author} {\bibfnamefont {J.~L.}\ \bibnamefont
  {Rosa}}\ and\ \bibinfo {author} {\bibfnamefont {D.}~\bibnamefont
  {Rubiera-Garcia}},\ }\href {http://dx.doi.org/10.1103/PhysRevD.106.084004}
  {\bibfield  {journal} {\bibinfo  {journal} {Physical Review D}\ }\textbf
  {\bibinfo {volume} {106}} (\bibinfo {year} {2022})}\BibitemShut {NoStop}%
\bibitem [{\citenamefont {Sanchis-Gual}\ \emph {et~al.}(2019)\citenamefont
  {Sanchis-Gual}, \citenamefont {Herdeiro}, \citenamefont {Font}, \citenamefont
  {Radu},\ and\ \citenamefont {Di~Giovanni}}]{Sanchis-Gual_2019}%
  \BibitemOpen
  \bibfield  {author} {\bibinfo {author} {\bibfnamefont {N.}~\bibnamefont
  {Sanchis-Gual}}, \bibinfo {author} {\bibfnamefont {C.}~\bibnamefont
  {Herdeiro}}, \bibinfo {author} {\bibfnamefont {J.~A.}\ \bibnamefont {Font}},
  \bibinfo {author} {\bibfnamefont {E.}~\bibnamefont {Radu}}, \ and\ \bibinfo
  {author} {\bibfnamefont {F.}~\bibnamefont {Di~Giovanni}},\ }\href
  {http://dx.doi.org/10.1103/PhysRevD.99.024017} {\bibfield  {journal}
  {\bibinfo  {journal} {Physical Review D}\ }\textbf {\bibinfo {volume} {99}}
  (\bibinfo {year} {2019})}\BibitemShut {NoStop}%
\bibitem [{\citenamefont {Bustillo}\ \emph {et~al.}(2023)\citenamefont
  {Bustillo}, \citenamefont {Sanchis-Gual}, \citenamefont {Leong},
  \citenamefont {Chandra}, \citenamefont {Torres-Forne}, \citenamefont {Font},
  \citenamefont {Herdeiro}, \citenamefont {Radu}, \citenamefont {Wong},\ and\
  \citenamefont {Li}}]{Bustillo_2023}%
  \BibitemOpen
  \bibfield  {author} {\bibinfo {author} {\bibfnamefont {J.~C.}\ \bibnamefont
  {Bustillo}}, \bibinfo {author} {\bibfnamefont {N.}~\bibnamefont
  {Sanchis-Gual}}, \bibinfo {author} {\bibfnamefont {S.~H.~W.}\ \bibnamefont
  {Leong}}, \bibinfo {author} {\bibfnamefont {K.}~\bibnamefont {Chandra}},
  \bibinfo {author} {\bibfnamefont {A.}~\bibnamefont {Torres-Forne}}, \bibinfo
  {author} {\bibfnamefont {J.~A.}\ \bibnamefont {Font}}, \bibinfo {author}
  {\bibfnamefont {C.}~\bibnamefont {Herdeiro}}, \bibinfo {author}
  {\bibfnamefont {E.}~\bibnamefont {Radu}}, \bibinfo {author} {\bibfnamefont
  {I.~C.~F.}\ \bibnamefont {Wong}}, \ and\ \bibinfo {author} {\bibfnamefont
  {T.~G.~F.}\ \bibnamefont {Li}},\ }\href {https://arxiv.org/abs/2206.02551}
  {\enquote {\bibinfo {title} {Searching for vector boson-star mergers within
  ligo-virgo intermediate-mass black-hole merger candidates},}\ } (\bibinfo
  {year} {2023}),\ \Eprint {http://arxiv.org/abs/2206.02551} {2206.02551
  [gr-qc]} \BibitemShut {NoStop}%
\bibitem [{\citenamefont {Brito}\ \emph
  {et~al.}(2016{\natexlab{b}})\citenamefont {Brito}, \citenamefont {Cardoso},
  \citenamefont {Macedo}, \citenamefont {Okawa},\ and\ \citenamefont
  {Palenzuela}}]{Brito:2015yfh}%
  \BibitemOpen
  \bibfield  {author} {\bibinfo {author} {\bibfnamefont {R.}~\bibnamefont
  {Brito}}, \bibinfo {author} {\bibfnamefont {V.}~\bibnamefont {Cardoso}},
  \bibinfo {author} {\bibfnamefont {C.~F.~B.}\ \bibnamefont {Macedo}}, \bibinfo
  {author} {\bibfnamefont {H.}~\bibnamefont {Okawa}}, \ and\ \bibinfo {author}
  {\bibfnamefont {C.}~\bibnamefont {Palenzuela}},\ }\href {\doibase
  10.1103/PhysRevD.93.044045} {\bibfield  {journal} {\bibinfo  {journal} {Phys.
  Rev. D}\ }\textbf {\bibinfo {volume} {93}},\ \bibinfo {pages} {044045}
  (\bibinfo {year} {2016}{\natexlab{b}})},\ \Eprint
  {http://arxiv.org/abs/1512.00466} {1512.00466} \BibitemShut {NoStop}%
\bibitem [{\citenamefont {Visinelli}(2021)}]{Visinelli:2021uve}%
  \BibitemOpen
  \bibfield  {author} {\bibinfo {author} {\bibfnamefont {L.}~\bibnamefont
  {Visinelli}},\ }\href {\doibase 10.1142/S0218271821300068} {\bibfield
  {journal} {\bibinfo  {journal} {Int. J. Mod. Phys. D}\ }\textbf {\bibinfo
  {volume} {30}},\ \bibinfo {pages} {2130006} (\bibinfo {year} {2021})},\
  \Eprint {http://arxiv.org/abs/2109.05481} {arXiv:2109.05481 [gr-qc]}
  \BibitemShut {NoStop}%
\bibitem [{\citenamefont {Di~Giovanni}\ \emph {et~al.}(2018)\citenamefont
  {Di~Giovanni}, \citenamefont {Sanchis-Gual}, \citenamefont {Herdeiro},\ and\
  \citenamefont {Font}}]{DiGiovanni:2018bvo}%
  \BibitemOpen
  \bibfield  {author} {\bibinfo {author} {\bibfnamefont {F.}~\bibnamefont
  {Di~Giovanni}}, \bibinfo {author} {\bibfnamefont {N.}~\bibnamefont
  {Sanchis-Gual}}, \bibinfo {author} {\bibfnamefont {C.~A.~R.}\ \bibnamefont
  {Herdeiro}}, \ and\ \bibinfo {author} {\bibfnamefont {J.~A.}\ \bibnamefont
  {Font}},\ }\href {\doibase 10.1103/PhysRevD.98.064044} {\bibfield  {journal}
  {\bibinfo  {journal} {Phys. Rev. D}\ }\textbf {\bibinfo {volume} {98}},\
  \bibinfo {pages} {064044} (\bibinfo {year} {2018})},\ \Eprint
  {http://arxiv.org/abs/1803.04802} {arXiv:1803.04802 [gr-qc]} \BibitemShut
  {NoStop}%
\bibitem [{\citenamefont {Sanchis-Gual}\ \emph {et~al.}(2017)\citenamefont
  {Sanchis-Gual}, \citenamefont {Herdeiro}, \citenamefont {Radu}, \citenamefont
  {Degollado},\ and\ \citenamefont {Font}}]{Sanchis-Gual:2017bhw}%
  \BibitemOpen
  \bibfield  {author} {\bibinfo {author} {\bibfnamefont {N.}~\bibnamefont
  {Sanchis-Gual}}, \bibinfo {author} {\bibfnamefont {C.}~\bibnamefont
  {Herdeiro}}, \bibinfo {author} {\bibfnamefont {E.}~\bibnamefont {Radu}},
  \bibinfo {author} {\bibfnamefont {J.~C.}\ \bibnamefont {Degollado}}, \ and\
  \bibinfo {author} {\bibfnamefont {J.~A.}\ \bibnamefont {Font}},\ }\href
  {\doibase 10.1103/PhysRevD.95.104028} {\bibfield  {journal} {\bibinfo
  {journal} {Phys. Rev. D}\ }\textbf {\bibinfo {volume} {95}},\ \bibinfo
  {pages} {104028} (\bibinfo {year} {2017})},\ \Eprint
  {http://arxiv.org/abs/1702.04532} {arXiv:1702.04532 [gr-qc]} \BibitemShut
  {NoStop}%
\bibitem [{\citenamefont {Zhang}\ \emph {et~al.}(2023)\citenamefont {Zhang},
  \citenamefont {Huang},\ and\ \citenamefont {Wang}}]{Zhang:2023rwc}%
  \BibitemOpen
  \bibfield  {author} {\bibinfo {author} {\bibfnamefont {R.}~\bibnamefont
  {Zhang}}, \bibinfo {author} {\bibfnamefont {L.-X.}\ \bibnamefont {Huang}}, \
  and\ \bibinfo {author} {\bibfnamefont {Y.-Q.}\ \bibnamefont {Wang}},\
  }\href@noop {} {\  (\bibinfo {year} {2023})},\ \Eprint
  {http://arxiv.org/abs/2312.15755} {arXiv:2312.15755 [gr-qc]} \BibitemShut
  {NoStop}%
\bibitem [{\citenamefont {Salazar~Landea}\ and\ \citenamefont
  {Garc\'\i{}a}(2016)}]{SalazarLandea:2016bys}%
  \BibitemOpen
  \bibfield  {author} {\bibinfo {author} {\bibfnamefont {I.}~\bibnamefont
  {Salazar~Landea}}\ and\ \bibinfo {author} {\bibfnamefont {F.}~\bibnamefont
  {Garc\'\i{}a}},\ }\href {\doibase 10.1103/PhysRevD.94.104006} {\bibfield
  {journal} {\bibinfo  {journal} {Phys. Rev. D}\ }\textbf {\bibinfo {volume}
  {94}},\ \bibinfo {pages} {104006} (\bibinfo {year} {2016})},\ \Eprint
  {http://arxiv.org/abs/1608.00011} {arXiv:1608.00011 [hep-th]} \BibitemShut
  {NoStop}%
\bibitem [{\citenamefont {Lazarte}\ and\ \citenamefont
  {Alcubierre}(2024)}]{Lazarte:2024jyr}%
  \BibitemOpen
  \bibfield  {author} {\bibinfo {author} {\bibfnamefont {C.}~\bibnamefont
  {Lazarte}}\ and\ \bibinfo {author} {\bibfnamefont {M.}~\bibnamefont
  {Alcubierre}},\ }\href@noop {} {\bibfield  {journal} {\bibinfo  {journal}
  {Class. Quant. Grav.}\ }\textbf {\bibinfo {volume} {41}},\ \bibinfo {pages}
  {135003} (\bibinfo {year} {2024})}\BibitemShut {NoStop}%
\bibitem [{\citenamefont {Balakrishna}\ \emph
  {et~al.}(1998{\natexlab{a}})\citenamefont {Balakrishna}, \citenamefont
  {Seidel},\ and\ \citenamefont {Suen}}]{Balakrishna_1998}%
  \BibitemOpen
  \bibfield  {author} {\bibinfo {author} {\bibfnamefont {J.}~\bibnamefont
  {Balakrishna}}, \bibinfo {author} {\bibfnamefont {E.}~\bibnamefont {Seidel}},
  \ and\ \bibinfo {author} {\bibfnamefont {W.-M.}\ \bibnamefont {Suen}},\
  }\href {http://dx.doi.org/10.1103/PhysRevD.58.104004} {\bibfield  {journal}
  {\bibinfo  {journal} {Physical Review D}\ }\textbf {\bibinfo {volume} {58}}
  (\bibinfo {year} {1998}{\natexlab{a}})}\BibitemShut {NoStop}%
\bibitem [{\citenamefont {Collodel}\ \emph {et~al.}(2017)\citenamefont
  {Collodel}, \citenamefont {Kleihaus},\ and\ \citenamefont
  {Kunz}}]{Collodel_2017}%
  \BibitemOpen
  \bibfield  {author} {\bibinfo {author} {\bibfnamefont {L.~G.}\ \bibnamefont
  {Collodel}}, \bibinfo {author} {\bibfnamefont {B.}~\bibnamefont {Kleihaus}},
  \ and\ \bibinfo {author} {\bibfnamefont {J.}~\bibnamefont {Kunz}},\ }\href
  {http://dx.doi.org/10.1103/PhysRevD.96.084066} {\bibfield  {journal}
  {\bibinfo  {journal} {Physical Review D}\ }\textbf {\bibinfo {volume} {96}}
  (\bibinfo {year} {2017})}\BibitemShut {NoStop}%
\bibitem [{\citenamefont {Brito}\ \emph {et~al.}(2023)\citenamefont {Brito},
  \citenamefont {Herdeiro}, \citenamefont {Radu}, \citenamefont
  {Sanchis-Gual},\ and\ \citenamefont {Zilh\~ao}}]{Brito:2023fwr}%
  \BibitemOpen
  \bibfield  {author} {\bibinfo {author} {\bibfnamefont {M.}~\bibnamefont
  {Brito}}, \bibinfo {author} {\bibfnamefont {C.}~\bibnamefont {Herdeiro}},
  \bibinfo {author} {\bibfnamefont {E.}~\bibnamefont {Radu}}, \bibinfo {author}
  {\bibfnamefont {N.}~\bibnamefont {Sanchis-Gual}}, \ and\ \bibinfo {author}
  {\bibfnamefont {M.}~\bibnamefont {Zilh\~ao}},\ }\href {\doibase
  10.1103/PhysRevD.107.084022} {\bibfield  {journal} {\bibinfo  {journal}
  {Phys. Rev. D}\ }\textbf {\bibinfo {volume} {107}},\ \bibinfo {pages}
  {084022} (\bibinfo {year} {2023})},\ \Eprint
  {http://arxiv.org/abs/2302.08900} {arXiv:2302.08900 [gr-qc]} \BibitemShut
  {NoStop}%
\bibitem [{\citenamefont {Herdeiro}\ \emph {et~al.}(2024)\citenamefont
  {Herdeiro}, \citenamefont {Radu}, \citenamefont {Sanchis-Gual}, \citenamefont
  {Santos},\ and\ \citenamefont {dos Santos Costa~Filho}}]{Herdeiro:2024a}%
  \BibitemOpen
  \bibfield  {author} {\bibinfo {author} {\bibfnamefont {C.}~\bibnamefont
  {Herdeiro}}, \bibinfo {author} {\bibfnamefont {E.}~\bibnamefont {Radu}},
  \bibinfo {author} {\bibfnamefont {N.}~\bibnamefont {Sanchis-Gual}}, \bibinfo
  {author} {\bibfnamefont {N.}~\bibnamefont {Santos}}, \ and\ \bibinfo {author}
  {\bibfnamefont {E.}~\bibnamefont {dos Santos Costa~Filho}},\ }\href@noop {}
  {\bibfield  {journal} {\bibinfo  {journal} {Physics Letters B}\ }\textbf
  {\bibinfo {volume} {852}},\ \bibinfo {pages} {138595} (\bibinfo {year}
  {2024})}\BibitemShut {NoStop}%
\bibitem [{\citenamefont {Alcubierre}(2008)}]{Alcubierre08a}%
  \BibitemOpen
  \bibfield  {author} {\bibinfo {author} {\bibfnamefont {M.}~\bibnamefont
  {Alcubierre}},\ }\href@noop {} {\emph {\bibinfo {title} {Introduction to
  $3+1$ Numerical Relativity}}}\ (\bibinfo  {publisher} {Oxford Univ. Press},\
  \bibinfo {address} {New York},\ \bibinfo {year} {2008})\BibitemShut {NoStop}%
\bibitem [{\citenamefont {Shibata}\ and\ \citenamefont
  {Nakamura}(1995)}]{Shibata95}%
  \BibitemOpen
  \bibfield  {author} {\bibinfo {author} {\bibfnamefont {M.}~\bibnamefont
  {Shibata}}\ and\ \bibinfo {author} {\bibfnamefont {T.}~\bibnamefont
  {Nakamura}},\ }\href@noop {} {\bibfield  {journal} {\bibinfo  {journal}
  {Phys. Rev.}\ }\textbf {\bibinfo {volume} {D52}},\ \bibinfo {pages} {5428}
  (\bibinfo {year} {1995})}\BibitemShut {NoStop}%
\bibitem [{\citenamefont {Baumgarte}\ and\ \citenamefont
  {Shapiro}(1998)}]{Baumgarte:1998te}%
  \BibitemOpen
  \bibfield  {author} {\bibinfo {author} {\bibfnamefont {T.~W.}\ \bibnamefont
  {Baumgarte}}\ and\ \bibinfo {author} {\bibfnamefont {S.~L.}\ \bibnamefont
  {Shapiro}},\ }\href@noop {} {\bibfield  {journal} {\bibinfo  {journal} {Phys.
  Rev.}\ }\textbf {\bibinfo {volume} {D59}},\ \bibinfo {pages} {024007}
  (\bibinfo {year} {1998})},\ \Eprint {http://arxiv.org/abs/gr-qc/9810065}
  {gr-qc/9810065} \BibitemShut {NoStop}%
\bibitem [{\citenamefont {Alcubierre}\ and\ \citenamefont
  {Mendez}(2011)}]{Alcubierre:2010is}%
  \BibitemOpen
  \bibfield  {author} {\bibinfo {author} {\bibfnamefont {M.}~\bibnamefont
  {Alcubierre}}\ and\ \bibinfo {author} {\bibfnamefont {M.~D.}\ \bibnamefont
  {Mendez}},\ }\href {\doibase 10.1007/s10714-011-1202-x} {\bibfield  {journal}
  {\bibinfo  {journal} {Gen.Rel.Grav.}\ }\textbf {\bibinfo {volume} {43}},\
  \bibinfo {pages} {2769} (\bibinfo {year} {2011})},\ \Eprint
  {http://arxiv.org/abs/1010.4013} {arXiv:1010.4013 [gr-qc]} \BibitemShut
  {NoStop}%
\bibitem [{\citenamefont {Balakrishna}\ \emph
  {et~al.}(1998{\natexlab{b}})\citenamefont {Balakrishna}, \citenamefont
  {Seidel},\ and\ \citenamefont {Suen}}]{Balakrishna:1997ej}%
  \BibitemOpen
  \bibfield  {author} {\bibinfo {author} {\bibfnamefont {J.}~\bibnamefont
  {Balakrishna}}, \bibinfo {author} {\bibfnamefont {E.}~\bibnamefont {Seidel}},
  \ and\ \bibinfo {author} {\bibfnamefont {W.-M.}\ \bibnamefont {Suen}},\
  }\href {\doibase 10.1103/PhysRevD.58.104004} {\bibfield  {journal} {\bibinfo
  {journal} {Phys. Rev.}\ }\textbf {\bibinfo {volume} {D58}},\ \bibinfo {pages}
  {104004} (\bibinfo {year} {1998}{\natexlab{b}})},\ \Eprint
  {http://arxiv.org/abs/gr-qc/9712064} {arXiv:gr-qc/9712064 [gr-qc]}
  \BibitemShut {NoStop}%
\bibitem [{\citenamefont {Guzman}(2004)}]{Guzman04}%
  \BibitemOpen
  \bibfield  {author} {\bibinfo {author} {\bibfnamefont {F.}~\bibnamefont
  {Guzman}},\ }\href {\doibase 10.1103/PhysRevD.70.044033} {\bibfield
  {journal} {\bibinfo  {journal} {Phys.Rev.}\ }\textbf {\bibinfo {volume}
  {D70}},\ \bibinfo {pages} {044033} (\bibinfo {year} {2004})},\ \Eprint
  {http://arxiv.org/abs/0407054} {arXiv:0407054 [gr-qc]} \BibitemShut {NoStop}%
\bibitem [{\citenamefont {Guzman}(2009)}]{Guzman09}%
  \BibitemOpen
  \bibfield  {author} {\bibinfo {author} {\bibfnamefont {F.}~\bibnamefont
  {Guzman}},\ }\href@noop {} {\bibfield  {journal} {\bibinfo  {journal}
  {Revista Mexicana de Fisica}\ }\textbf {\bibinfo {volume} {55}},\ \bibinfo
  {pages} {321} (\bibinfo {year} {2009})}\BibitemShut {NoStop}%
\bibitem [{\citenamefont {Seidel}\ and\ \citenamefont {Suen}(1990)}]{Seidel90}%
  \BibitemOpen
  \bibfield  {author} {\bibinfo {author} {\bibfnamefont {E.}~\bibnamefont
  {Seidel}}\ and\ \bibinfo {author} {\bibfnamefont {W.}~\bibnamefont {Suen}},\
  }\href@noop {} {\bibfield  {journal} {\bibinfo  {journal} {Phys. Rev.}\
  }\textbf {\bibinfo {volume} {D42}},\ \bibinfo {pages} {384} (\bibinfo {year}
  {1990})}\BibitemShut {NoStop}%
\bibitem [{\citenamefont {Loginov}(2015)}]{Loginov_2015}%
  \BibitemOpen
  \bibfield  {author} {\bibinfo {author} {\bibfnamefont {A.~Y.}\ \bibnamefont
  {Loginov}},\ }\href {https://link.aps.org/doi/10.1103/PhysRevD.91.105028}
  {\bibfield  {journal} {\bibinfo  {journal} {Phys. Rev. D}\ }\textbf {\bibinfo
  {volume} {91}},\ \bibinfo {pages} {105028} (\bibinfo {year}
  {2015})}\BibitemShut {NoStop}%
\bibitem [{\citenamefont {Arbona}\ \emph {et~al.}(1999)\citenamefont {Arbona},
  \citenamefont {Bona}, \citenamefont {Mass{\'o}},\ and\ \citenamefont
  {Stela}}]{Arbona99}%
  \BibitemOpen
  \bibfield  {author} {\bibinfo {author} {\bibfnamefont {A.}~\bibnamefont
  {Arbona}}, \bibinfo {author} {\bibfnamefont {C.}~\bibnamefont {Bona}},
  \bibinfo {author} {\bibfnamefont {J.}~\bibnamefont {Mass{\'o}}}, \ and\
  \bibinfo {author} {\bibfnamefont {J.}~\bibnamefont {Stela}},\ }\href@noop {}
  {\bibfield  {journal} {\bibinfo  {journal} {Phys. Rev.}\ }\textbf {\bibinfo
  {volume} {D60}},\ \bibinfo {pages} {104014} (\bibinfo {year} {1999})},\
  \Eprint {http://arxiv.org/abs/gr-qc/9902053} {gr-qc/9902053} \BibitemShut
  {NoStop}%
\bibitem [{\citenamefont {Alcubierre}\ \emph {et~al.}(2003)\citenamefont
  {Alcubierre}, \citenamefont {Br\"ugmann}, \citenamefont {Diener},
  \citenamefont {Koppitz}, \citenamefont {Pollney}, \citenamefont {Seidel},\
  and\ \citenamefont {Takahashi}}]{Alcubierre02a}%
  \BibitemOpen
  \bibfield  {author} {\bibinfo {author} {\bibfnamefont {M.}~\bibnamefont
  {Alcubierre}}, \bibinfo {author} {\bibfnamefont {B.}~\bibnamefont
  {Br\"ugmann}}, \bibinfo {author} {\bibfnamefont {P.}~\bibnamefont {Diener}},
  \bibinfo {author} {\bibfnamefont {M.}~\bibnamefont {Koppitz}}, \bibinfo
  {author} {\bibfnamefont {D.}~\bibnamefont {Pollney}}, \bibinfo {author}
  {\bibfnamefont {E.}~\bibnamefont {Seidel}}, \ and\ \bibinfo {author}
  {\bibfnamefont {R.}~\bibnamefont {Takahashi}},\ }\href@noop {} {\bibfield
  {journal} {\bibinfo  {journal} {Phys. Rev.}\ }\textbf {\bibinfo {volume}
  {D67}},\ \bibinfo {pages} {084023} (\bibinfo {year} {2003})},\ \Eprint
  {http://arxiv.org/abs/gr-qc/0206072} {gr-qc/0206072} \BibitemShut {NoStop}%
\bibitem [{\citenamefont {Alcubierre}\ \emph {et~al.}(2001)\citenamefont
  {Alcubierre}, \citenamefont {Br\"ugmann}, \citenamefont {Pollney},
  \citenamefont {Seidel},\ and\ \citenamefont {Takahashi}}]{Alcubierre01a}%
  \BibitemOpen
  \bibfield  {author} {\bibinfo {author} {\bibfnamefont {M.}~\bibnamefont
  {Alcubierre}}, \bibinfo {author} {\bibfnamefont {B.}~\bibnamefont
  {Br\"ugmann}}, \bibinfo {author} {\bibfnamefont {D.}~\bibnamefont {Pollney}},
  \bibinfo {author} {\bibfnamefont {E.}~\bibnamefont {Seidel}}, \ and\ \bibinfo
  {author} {\bibfnamefont {R.}~\bibnamefont {Takahashi}},\ }\href@noop {}
  {\bibfield  {journal} {\bibinfo  {journal} {Phys. Rev.}\ }\textbf {\bibinfo
  {volume} {D64}},\ \bibinfo {pages} {R61501} (\bibinfo {year} {2001})},\
  \Eprint {http://arxiv.org/abs/gr-qc/0104020} {gr-qc/0104020} \BibitemShut
  {NoStop}%
\bibitem [{\citenamefont {Alcubierre}\ and\ \citenamefont
  {Torres}(2015)}]{Alcubierre:2014joa}%
  \BibitemOpen
  \bibfield  {author} {\bibinfo {author} {\bibfnamefont {M.}~\bibnamefont
  {Alcubierre}}\ and\ \bibinfo {author} {\bibfnamefont {J.~M.}\ \bibnamefont
  {Torres}},\ }\href {\doibase 10.1088/0264-9381/32/3/035006} {\bibfield
  {journal} {\bibinfo  {journal} {Class. Quant. Grav.}\ }\textbf {\bibinfo
  {volume} {32}},\ \bibinfo {pages} {035006} (\bibinfo {year} {2015})},\
  \Eprint {http://arxiv.org/abs/1407.8529} {arXiv:1407.8529 [gr-qc]}
  \BibitemShut {NoStop}%
\bibitem [{\citenamefont {Alcubierre}\ \emph {et~al.}(2019)\citenamefont
  {Alcubierre}, \citenamefont {Barranco}, \citenamefont {Bernal}, \citenamefont
  {Degollado}, \citenamefont {Diez-Tejedor}, \citenamefont {Megevand},
  \citenamefont {N\'u\~nez},\ and\ \citenamefont
  {Sarbach}}]{Alcubierre:2019qnh}%
  \BibitemOpen
  \bibfield  {author} {\bibinfo {author} {\bibfnamefont {M.}~\bibnamefont
  {Alcubierre}}, \bibinfo {author} {\bibfnamefont {J.}~\bibnamefont
  {Barranco}}, \bibinfo {author} {\bibfnamefont {A.}~\bibnamefont {Bernal}},
  \bibinfo {author} {\bibfnamefont {J.~C.}\ \bibnamefont {Degollado}}, \bibinfo
  {author} {\bibfnamefont {A.}~\bibnamefont {Diez-Tejedor}}, \bibinfo {author}
  {\bibfnamefont {M.}~\bibnamefont {Megevand}}, \bibinfo {author}
  {\bibfnamefont {D.}~\bibnamefont {N\'u\~nez}}, \ and\ \bibinfo {author}
  {\bibfnamefont {O.}~\bibnamefont {Sarbach}},\ }\href {\doibase
  10.1088/1361-6382/ab4726} {\bibfield  {journal} {\bibinfo  {journal} {Class.
  Quant. Grav.}\ }\textbf {\bibinfo {volume} {36}},\ \bibinfo {pages} {215013}
  (\bibinfo {year} {2019})},\ \Eprint {http://arxiv.org/abs/1906.08959}
  {arXiv:1906.08959 [gr-qc]} \BibitemShut {NoStop}%
\bibitem [{\citenamefont {Degollado}\ \emph {et~al.}(2020)\citenamefont
  {Degollado}, \citenamefont {Salgado},\ and\ \citenamefont
  {Alcubierre}}]{Degollado:2020lsa}%
  \BibitemOpen
  \bibfield  {author} {\bibinfo {author} {\bibfnamefont {J.~C.}\ \bibnamefont
  {Degollado}}, \bibinfo {author} {\bibfnamefont {M.}~\bibnamefont {Salgado}},
  \ and\ \bibinfo {author} {\bibfnamefont {M.}~\bibnamefont {Alcubierre}},\
  }\href {\doibase 10.1016/j.physletb.2020.135666} {\bibfield  {journal}
  {\bibinfo  {journal} {Phys. Lett. B}\ }\textbf {\bibinfo {volume} {808}},\
  \bibinfo {pages} {135666} (\bibinfo {year} {2020})},\ \Eprint
  {http://arxiv.org/abs/2008.10683} {arXiv:2008.10683 [gr-qc]} \BibitemShut
  {NoStop}%
\bibitem [{\citenamefont {Jiménez-Vázquez}\ and\ \citenamefont
  {Alcubierre}(2022{\natexlab{a}})}]{Jimenez2022a}%
  \BibitemOpen
  \bibfield  {author} {\bibinfo {author} {\bibfnamefont {E.}~\bibnamefont
  {Jiménez-Vázquez}}\ and\ \bibinfo {author} {\bibfnamefont {M.}~\bibnamefont
  {Alcubierre}},\ }\href {\doibase 10.1103/physrevd.105.064071} {\bibfield
  {journal} {\bibinfo  {journal} {Physical Review D}\ }\textbf {\bibinfo
  {volume} {105}} (\bibinfo {year} {2022}{\natexlab{a}}),\
  10.1103/physrevd.105.064071}\BibitemShut {NoStop}%
\bibitem [{\citenamefont {Jiménez-Vázquez}\ and\ \citenamefont
  {Alcubierre}(2022{\natexlab{b}})}]{Jimenez2022b}%
  \BibitemOpen
  \bibfield  {author} {\bibinfo {author} {\bibfnamefont {E.}~\bibnamefont
  {Jiménez-Vázquez}}\ and\ \bibinfo {author} {\bibfnamefont {M.}~\bibnamefont
  {Alcubierre}},\ }\href {\doibase 10.1103/physrevd.106.044071} {\bibfield
  {journal} {\bibinfo  {journal} {Physical Review D}\ }\textbf {\bibinfo
  {volume} {106}} (\bibinfo {year} {2022}{\natexlab{b}}),\
  10.1103/physrevd.106.044071}\BibitemShut {NoStop}%
\bibitem [{\citenamefont {Brito}\ \emph
  {et~al.}(2016{\natexlab{c}})\citenamefont {Brito}, \citenamefont {Cardoso},
  \citenamefont {Herdeiro},\ and\ \citenamefont {Radu}}]{Brito:2015pxa}%
  \BibitemOpen
  \bibfield  {author} {\bibinfo {author} {\bibfnamefont {R.}~\bibnamefont
  {Brito}}, \bibinfo {author} {\bibfnamefont {V.}~\bibnamefont {Cardoso}},
  \bibinfo {author} {\bibfnamefont {C.~A.~R.}\ \bibnamefont {Herdeiro}}, \ and\
  \bibinfo {author} {\bibfnamefont {E.}~\bibnamefont {Radu}},\ }\href {\doibase
  10.1016/j.physletb.2015.11.051} {\bibfield  {journal} {\bibinfo  {journal}
  {Phys. Lett. B}\ }\textbf {\bibinfo {volume} {752}},\ \bibinfo {pages} {291}
  (\bibinfo {year} {2016}{\natexlab{c}})},\ \Eprint
  {http://arxiv.org/abs/1508.05395} {arXiv:1508.05395 [gr-qc]} \BibitemShut
  {NoStop}%
\bibitem [{\citenamefont {Herdeiro}\ \emph {et~al.}(2017)\citenamefont
  {Herdeiro}, \citenamefont {Pombo},\ and\ \citenamefont
  {Radu}}]{Herdeiro:2017fhv}%
  \BibitemOpen
  \bibfield  {author} {\bibinfo {author} {\bibfnamefont {C.~A.~R.}\
  \bibnamefont {Herdeiro}}, \bibinfo {author} {\bibfnamefont {A.~M.}\
  \bibnamefont {Pombo}}, \ and\ \bibinfo {author} {\bibfnamefont
  {E.}~\bibnamefont {Radu}},\ }\href {\doibase 10.1016/j.physletb.2017.09.036}
  {\bibfield  {journal} {\bibinfo  {journal} {Phys. Lett. B}\ }\textbf
  {\bibinfo {volume} {773}},\ \bibinfo {pages} {654} (\bibinfo {year}
  {2017})},\ \Eprint {http://arxiv.org/abs/1708.05674} {arXiv:1708.05674
  [gr-qc]} \BibitemShut {NoStop}%
\bibitem [{\citenamefont {Cardoso}\ and\ \citenamefont
  {Pani}(2019)}]{Cardoso:2019rvt}%
  \BibitemOpen
  \bibfield  {author} {\bibinfo {author} {\bibfnamefont {V.}~\bibnamefont
  {Cardoso}}\ and\ \bibinfo {author} {\bibfnamefont {P.}~\bibnamefont {Pani}},\
  }\href {\doibase 10.1007/s41114-019-0020-4} {\bibfield  {journal} {\bibinfo
  {journal} {Living Rev. Rel.}\ }\textbf {\bibinfo {volume} {22}},\ \bibinfo
  {pages} {4} (\bibinfo {year} {2019})},\ \Eprint
  {http://arxiv.org/abs/1904.05363} {arXiv:1904.05363 [gr-qc]} \BibitemShut
  {NoStop}%
\bibitem [{\citenamefont {Sanchis-Gual}\ \emph {et~al.}(2022)\citenamefont
  {Sanchis-Gual}, \citenamefont {Herdeiro},\ and\ \citenamefont
  {Radu}}]{Sanchis_Gual_2022}%
  \BibitemOpen
  \bibfield  {author} {\bibinfo {author} {\bibfnamefont {N.}~\bibnamefont
  {Sanchis-Gual}}, \bibinfo {author} {\bibfnamefont {C.}~\bibnamefont
  {Herdeiro}}, \ and\ \bibinfo {author} {\bibfnamefont {E.}~\bibnamefont
  {Radu}},\ }\href {http://dx.doi.org/10.1088/1361-6382/ac4b9b} {\bibfield
  {journal} {\bibinfo  {journal} {Classical and Quantum Gravity}\ }\textbf
  {\bibinfo {volume} {39}},\ \bibinfo {pages} {064001} (\bibinfo {year}
  {2022})}\BibitemShut {NoStop}%
\bibitem [{\citenamefont {Di~Giovanni}\ \emph {et~al.}(2021)\citenamefont
  {Di~Giovanni}, \citenamefont {Fakhry}, \citenamefont {Sanchis-Gual},
  \citenamefont {Degollado},\ and\ \citenamefont {Font}}]{DiGiovanni:2021vlu}%
  \BibitemOpen
  \bibfield  {author} {\bibinfo {author} {\bibfnamefont {F.}~\bibnamefont
  {Di~Giovanni}}, \bibinfo {author} {\bibfnamefont {S.}~\bibnamefont {Fakhry}},
  \bibinfo {author} {\bibfnamefont {N.}~\bibnamefont {Sanchis-Gual}}, \bibinfo
  {author} {\bibfnamefont {J.~C.}\ \bibnamefont {Degollado}}, \ and\ \bibinfo
  {author} {\bibfnamefont {J.~A.}\ \bibnamefont {Font}},\ }\href {\doibase
  10.1088/1361-6382/ac1b45} {\bibfield  {journal} {\bibinfo  {journal} {Class.
  Quant. Grav.}\ }\textbf {\bibinfo {volume} {38}},\ \bibinfo {pages} {194001}
  (\bibinfo {year} {2021})},\ \Eprint {http://arxiv.org/abs/2105.00530}
  {arXiv:2105.00530 [gr-qc]} \BibitemShut {NoStop}%
\bibitem [{\citenamefont {Clough}\ \emph {et~al.}(2022)\citenamefont {Clough},
  \citenamefont {Helfer}, \citenamefont {Witek},\ and\ \citenamefont
  {Berti}}]{Clough_2022}%
  \BibitemOpen
  \bibfield  {author} {\bibinfo {author} {\bibfnamefont {K.}~\bibnamefont
  {Clough}}, \bibinfo {author} {\bibfnamefont {T.}~\bibnamefont {Helfer}},
  \bibinfo {author} {\bibfnamefont {H.}~\bibnamefont {Witek}}, \ and\ \bibinfo
  {author} {\bibfnamefont {E.}~\bibnamefont {Berti}},\ }\href
  {http://dx.doi.org/10.1103/PhysRevLett.129.151102} {\bibfield  {journal}
  {\bibinfo  {journal} {Physical Review Letters}\ }\textbf {\bibinfo {volume}
  {129}} (\bibinfo {year} {2022})}\BibitemShut {NoStop}%
\bibitem [{\citenamefont {Coates}\ and\ \citenamefont
  {Ramazanoğlu}(2022)}]{Coates_2022}%
  \BibitemOpen
  \bibfield  {author} {\bibinfo {author} {\bibfnamefont {A.}~\bibnamefont
  {Coates}}\ and\ \bibinfo {author} {\bibfnamefont {F.~M.}\ \bibnamefont
  {Ramazanoğlu}},\ }\href {http://dx.doi.org/10.1103/PhysRevLett.129.151103}
  {\bibfield  {journal} {\bibinfo  {journal} {Physical Review Letters}\
  }\textbf {\bibinfo {volume} {129}} (\bibinfo {year} {2022})}\BibitemShut
  {NoStop}%
\bibitem [{\citenamefont {Aoki}\ and\ \citenamefont
  {Minamitsuji}(2022)}]{Aoki:2022woy}%
  \BibitemOpen
  \bibfield  {author} {\bibinfo {author} {\bibfnamefont {K.}~\bibnamefont
  {Aoki}}\ and\ \bibinfo {author} {\bibfnamefont {M.}~\bibnamefont
  {Minamitsuji}},\ }\href {\doibase 10.1103/PhysRevD.106.084022} {\bibfield
  {journal} {\bibinfo  {journal} {Phys. Rev. D}\ }\textbf {\bibinfo {volume}
  {106}},\ \bibinfo {pages} {084022} (\bibinfo {year} {2022})},\ \Eprint
  {http://arxiv.org/abs/2206.14320} {arXiv:2206.14320 [gr-qc]} \BibitemShut
  {NoStop}%
\bibitem [{\citenamefont {Herdeiro}\ \emph {et~al.}(2023)\citenamefont
  {Herdeiro}, \citenamefont {Radu},\ and\ \citenamefont {dos Santos
  Costa~Filho}}]{Herdeiro:2023a}%
  \BibitemOpen
  \bibfield  {author} {\bibinfo {author} {\bibfnamefont {C.}~\bibnamefont
  {Herdeiro}}, \bibinfo {author} {\bibfnamefont {E.}~\bibnamefont {Radu}}, \
  and\ \bibinfo {author} {\bibfnamefont {E.}~\bibnamefont {dos Santos
  Costa~Filho}},\ }\href {http://dx.doi.org/10.1088/1475-7516/2023/05/022}
  {\bibfield  {journal} {\bibinfo  {journal} {Journal of Cosmology and
  Astroparticle Physics}\ }\textbf {\bibinfo {volume} {2023}},\ \bibinfo
  {pages} {022} (\bibinfo {year} {2023})}\BibitemShut {NoStop}%
\bibitem [{\citenamefont {Brito}\ \emph {et~al.}(2024)\citenamefont {Brito},
  \citenamefont {Herdeiro}, \citenamefont {Sanchis-Gual}, \citenamefont {dos
  Santos Costa~Filho},\ and\ \citenamefont {Zilhao}}]{Brito:2024}%
  \BibitemOpen
  \bibfield  {author} {\bibinfo {author} {\bibfnamefont {M.}~\bibnamefont
  {Brito}}, \bibinfo {author} {\bibfnamefont {C.}~\bibnamefont {Herdeiro}},
  \bibinfo {author} {\bibfnamefont {N.}~\bibnamefont {Sanchis-Gual}}, \bibinfo
  {author} {\bibfnamefont {E.}~\bibnamefont {dos Santos Costa~Filho}}, \ and\
  \bibinfo {author} {\bibfnamefont {M.}~\bibnamefont {Zilhao}},\ }\href@noop {}
  {\bibfield  {journal} {\bibinfo  {journal} {Classical and Quantum Gravity}\
  }\textbf {\bibinfo {volume} {41}},\ \bibinfo {pages} {195005} (\bibinfo
  {year} {2024})}\BibitemShut {NoStop}%
\bibitem [{\citenamefont {Rubio}\ \emph {et~al.}(2024)\citenamefont {Rubio},
  \citenamefont {Lara}, \citenamefont {Bezares}, \citenamefont {Crisostomi},\
  and\ \citenamefont {Barausse}}]{Rubio:2024ryv}%
  \BibitemOpen
  \bibfield  {author} {\bibinfo {author} {\bibfnamefont {M.~E.}\ \bibnamefont
  {Rubio}}, \bibinfo {author} {\bibfnamefont {G.}~\bibnamefont {Lara}},
  \bibinfo {author} {\bibfnamefont {M.}~\bibnamefont {Bezares}}, \bibinfo
  {author} {\bibfnamefont {M.}~\bibnamefont {Crisostomi}}, \ and\ \bibinfo
  {author} {\bibfnamefont {E.}~\bibnamefont {Barausse}},\ }\href@noop {}
  {\bibfield  {journal} {\bibinfo  {journal} {Phys. Rev. D}\ }\textbf {\bibinfo
  {volume} {110}},\ \bibinfo {pages} {063015} (\bibinfo {year}
  {2024})}\BibitemShut {NoStop}%
\end{thebibliography}%


\end{document}